

%
%
%
\def\and{{\it\&}}
\def\half{{1\over2}}
\def\third{{1\over3}}

\def\gesim{\,{\raise-3pt\hbox{$\sim$}}\!\!\!\!\!{\raise2pt\hbox{$>$}}\,}
\def\lesim{\,{\raise-3pt\hbox{$\sim$}}\!\!\!\!\!{\raise2pt\hbox{$<$}}\,}
\def\boldoverdot{\,{\raise6pt\hbox{\bf.}\!\!\!\!\>}}

\def\ie{{\it i.e.}}

\def\ibid{{\it ibid.}}
\def\etal{{\it et. al.}}
\def\acal{{\cal A}}

\def\ccal{{\cal C}}
\def\dcal{{\cal D}}

\def\jcal{{\cal J}}

\def\lcal{{\cal L}}

\def\ocal{{\cal O}}
\def\pcal{{\cal P}}

\def\tcal{{\cal T}}

\def\WW{{\bf W}}

\def\ZZ{{\bf Z}}

\def\lBB{ \hbox{{\smallii I}}\!\hbox{{\smallii L}} }

\def\vev{vacuum expectation value}

\def\tr{ \hbox{tr}}

\def\diag{\hbox{\diag}}
\def\sm{Standard Model}

\def\gev{~\hbox{GeV}}
\def\tev{~\hbox{TeV}}

\def\kg{~\hbox{kg}}
%
%
%
\def\leaderfill{\leaders\hbox to 1 em{\hss.\hss}\hfill}
\def\inbox#1{\vbox{\hrule\hbox{\vrule\kern5pt
     \vbox{\kern5pt#1\kern5pt}\kern5pt\vrule}\hrule}}
\def\sqr#1#2{{\vcenter{\hrule height.#2pt
      \hbox{\vrule width.#2pt height#1pt \kern#1pt
         \vrule width.#2pt}
      \hrule height.#2pt}}}
\def\square{\mathchoice\sqr56\sqr56\sqr{2.1}3\sqr{1.5}3}
\def\today{\ifcase\month\or
  January\or February\or March\or April\or May\or June\or
  July\or August\or September\or October\or November\or December\fi
  \space\number\day, \number\year}
\def\pmb#1{\setbox0=\hbox{#1}%
  \kern-.025em\copy0\kern-\wd0
  \kern.05em\copy0\kern-\wd0
  \kern-.025em\raise.0433em\box0 }
\def\up#1{^{\left( #1 \right) }}
\def\lowti#1{_{{\rm #1 }}}
\def\inv#1{{1\over#1}}
\def\su#1{{SU(#1)}}
\def\ui{U(1)}
\def\antes{}
\def\despues{.}
%

%
\def\sumprime_#1{\setbox0=\hbox{$\scriptstyle{#1}$}
  \setbox2=\hbox{$\displaystyle{\sum}$}
  \setbox4=\hbox{${}'\mathsurround=0pt$}
  \dimen0=.5\wd0 \advance\dimen0 by-.5\wd2
  \ifdim\dimen0>0pt
  \ifdim\dimen0>\wd4 \kern\wd4 \else\kern\dimen0\fi\fi
\mathop{{\sum}'}_{\kern-\wd4 #1}}
\def\sumbiprime_#1{\setbox0=\hbox{$\scriptstyle{#1}$}
  \setbox2=\hbox{$\displaystyle{\sum}$}
  \setbox4=\hbox{${}'\mathsurround=0pt$}
  \dimen0=.5\wd0 \advance\dimen0 by-.5\wd2
  \ifdim\dimen0>0pt
  \ifdim\dimen0>\wd4 \kern\wd4 \else\kern\dimen0\fi\fi
\mathop{{\sum}''}_{\kern-\wd4 #1}}
\def\sumtriprime_#1{\setbox0=\hbox{$\scriptstyle{#1}$}
  \setbox2=\hbox{$\displaystyle{\sum}$}
  \setbox4=\hbox{${}'\mathsurround=0pt$}
  \dimen0=.5\wd0 \advance\dimen0 by-.5\wd2
  \ifdim\dimen0>0pt
  \ifdim\dimen0>\wd4 \kern\wd4 \else\kern\dimen0\fi\fi
\mathop{{\sum}'''}_{\kern-\wd4 #1}}
%
%
\newcount\chapnum
\def\clearchap{\chapnum=0}
\def\chap#1{\clearsect\clearprob
\global\advance\chapnum by 1 \par\vskip .5 in\par%
\centerline{{\bigboldiii\antes\the\chapnum\despues\ #1}}}
\newcount\sectnum
\def\clearsect{\sectnum=0}
\def\sect#1{\clearprob\global\advance\sectnum by 1 \par\vskip .25 in\par%
\noindent{\bigboldii\the\chapnum.\the\sectnum:\ #1}\nobreak}
\newcount\yesnonum
\def\clearyesno{\yesnonum=0}
\def\verify{\global\advance\yesnonum by 1{\bigboldi (VERIFY!!)}}
\def\tocheck{\par\vskip 1 in{\bigboldv TO VERIFY: \the\yesnonum\ ITEMS.}}
\newcount\notenum

\def\note#1{\global\advance\notenum by 1{ \bf $<<$ #1 $>>$ } }
\def\noteout{\par\vskip 1 in{\bigboldiv NOTES: \the\notenum.}}
\newcount\borrownum

\def\borrow{\global\advance\borrownum by 1{\bigboldi BORROWED BY:\ }}
\def\borrowed{\par\vskip 0.5 in{\bigboldii BOOKS OUT:\ \the\borrownum.}}
\newcount\refnum

\def\ref#1{\global\advance\refnum by 1\item{\the\refnum.\ }#1}
\def\stariref#1{\global\advance\refnum by 1\item{%
               {\bigboldiv *}\the\refnum.\ }#1}
\def\stariiref#1{\global\advance\refnum by 1\item{%
               {\bigboldiv **}\the\refnum.\ }#1}
\def\stariiiref#1{\global\advance\refnum by 1\item{
               {\bigboldiv ***}\the\refnum.\ }#1}
\newcount\probnum
\def\clearprob{\probnum=0}
\def\prob{\global\advance\probnum by 1 {\medskip $\triangleright$\
\undertext{{\sl Problem}}\ \the\chapnum.\the\sectnum.\the\probnum.\ }}
\newcount\probchapnum

\def\probchap{\global\advance\probchapnum by 1 {\medskip $\triangleright$\
\undertext{{\sl Problem}}\ \the\chapnum.\the\probchapnum.\ }}
\def\undertext#1{$\underline{\smash{\hbox{#1}}}$}
%
%
%

\def\UCR{
{{\it University of California at Riverside\break
                  Department of Physics\break
                  Riverside, California 92521--0413; U.S.A. \break
                  WUDKA{\rm @}UCRPHYS}}}

%
%
%
%
\font\sanser=cmssq8

%

%

%
\font\bigboldi=cmbx10 scaled\magstep1
\font\bigboldii=cmbx10 scaled\magstep2
\font\bigboldiii=cmbx10 scaled\magstep3
\font\bigboldiv=cmbx10 scaled\magstep4
\font\bigboldv=cmbx10 scaled\magstep5
\font\small=cmr8
\font\smalli=cmr8 scaled\magstep1
\font\smallii=cmr8 scaled\magstep2

\font\eightrm=cmr8
%
%
%
%
\clearchap
\clearyesno
\headline={\ifnum\pageno>0   {\smalli \title (\today)} \hfil {\small Page
\folio } \else\hfil\fi}
%
%
%
%
\newdimen\fullhsize
\newdimen\fullvsize
\newbox\leftcolumn
\def\fullline{\hbox to\fullhsize}
\gdef\twocol{\fullhsize=9.75in
\hsize=4.6in
\vsize=7in
\advance\hoffset by -.5 in
\def\makeheadline{\vbox to 0pt{\vskip-.4in
  \fullline{\vbox to8.5pt{}\the\headline}\vss}
   \nointerlineskip}
\def\makefootline{\baselineskip=24pt
    \fullline{\the\footline}}
\let\lr=L
\output{\if L\lr
   \global\setbox\leftcolumn=\columnbox \global\let\lr=R
  \else \doubleformat \global\let\lr=L\fi
        \ifnum\outputpenalty>-2000 \else\dosupereject\fi}
\def\doubleformat{\shipout\vbox{\makeheadline
     \fullline{\box\leftcolumn\hfil\columnbox}\makefootline
     }\advancepageno}
\def\columnbox{\leftline{\pagebody}}
\nopagenumbers
\hfuzz=3pt}
%
%
%
%
%
%

%
%

%


\input epsf.tex

\def\title{Course on effective larangians: for camera-ready publication}

\def\theabstract{{\tenrm
In this set of lectures I will present an elementary
introduction to the uses of effective lagrangians in the electroweak
interaction with emphasis in practical applications}}

\def\thetitle{{\bf EFFECTIVE LAGRANGIANS (for electroweak physics)}
\footnote*{Lectures presented at the
{\sl IV Workshop on Particles and
Fields of the Mexican Physical Society.} M\'erida, Yucat\'an, M\'exico,
Oct. 1993.}}
\def\ucrnum{121}

\newlinechar=`\^^J
\immediate\write16{^^J ONE COLUMN OUTPUT  ^^J}

\input phyzzx
\PHYSREV
\rightline{UCRHEP-T\ucrnum}
{\titlepage
\vskip -.2 in
\title{ {\bigboldiii \thetitle}}
\doublespace
\author{{ Jos\'e Wudka }}
\address{\UCR}
\abstract
\bigskip
\singlespace
\theabstract
\endpage}
\hfuzz 34 pt
\singlespace


\def\square{\mathchoice\sqr56\sqr56\sqr{2.1}3\sqr{1.5}3}
\def\vev{vacuum expectation value}
\def\cuv{{C_{UV}}}
\def\leff{\lcal\lowti{eff}}
\def\mw{m\lowti w}
\def\sw{s\lowti w}
\def\cw{c\lowti w}
\def\tw{t\lowti w}
\def\lb{\Lambda_\phi}
\def\lf{\Lambda_\psi}
\def\ibos{{I_b}}
\def\ifer{{I_f}}
\def\mp{m\lowti{phys}}
\def\lp{\lambda\lowti{phys}}
\def\ap{\alpha\lowti{phys}}
\def\plb{{\it Phys. Lett.} {\bf B}}

\def\prl{{\it Phys. Rev. Lett.}}
\def\prd{{\it Phys. Rev.} {\bf D}}
\def\npb{{\it Nucl. Phys.} {\bf B}}
\def\zphys{{\it Z. Phys.} {\bf C}}
\def\Lambdauv{\Lambda_{UV}}

\chapter{Introduction}


I will describe the use of effective lagrangians and
delineate the philosophy on which the method is
based. I will then review the various theoretical details relevant to the
construction of effective lagrangians insuring consistency with know
experimental data.

Due to time limitations I will not be able to thoroughly cover the
subject. To repair this deficiency I will refer the reader to
the literature. Since this is intended to be a pedagogical
review I have referred, whenever possible,
to books and review articles instead of the
original papers. The complete set of references can be found in the
publications cited here.

\section{{\it Motivation.}}

\REF\itzykson{C. Itzykson and J.-B. Zuber, {\it Quantum Field Theory}
              (McGraw-Hill, New York, 1980).}
\REF\schrieffer{J.R. Schrieffer, {\it Theory of Superconductivity}
                (W.A. Benjamin, New York 1964).}
\REF\pich{A. Pich, lectures presented at the {\it V Mexican School of
          Particles and Fields}, Guanajuato, M\'exico, Dec. 1992.}
Since the very first applications of field theory a
very useful method in the description of certain phenomena has been
the use of effective lagrangians. Classical examples are the Fermi
theory of the weak interactions~\refmark{\itzykson} and the BCS theory of
superconductivity~\refmark{\schrieffer}.
A more recent example, equally successful, is the
use of chiral lagrangians in the description of the strong interactions
at low energies (for a review see
Ref.~\pich).

All of the above theories were based on a set of well established
experimental facts, such as the relevant particles involved in
the phenomena under consideration, and the symmetries respected
by them. It is understood
that these models do not provide the ultimate, or most
profound, description of the processes studied, but that
there is a fundamental theory which coincides with the
effective (phenomenological) model in the situations where the
latter is applicable. For example, the four-Fermi interactions
are now known to be the low energy remnants of the $W$ and $Z$
exchange interactions among fermions. An attractive feature of the
effective lagrangian method is the minimal number of assumptions
which are made concerning the underlying physics.

Another property of these theories is their limited range of applicability.
In fact, the Fermi
approach to the weak interactions is not to be applied above a certain
energy, the BCS theory was not intended to describe the behaviour of
metals at all temperatures, not are the chiral lagrangians a
substitute for QCD. All of these effective theories have a range of
validity, specified by a certain UV and IR scales, beyond which the
model lagrangian should not be blindly applied for it will generate
erroneous results. For example the Fermi current-current interaction
predictions do not match the experimental results for energies
comparable to the $W$ mass; the chiral description of the strong
interactions breaks down at energies of the order of $ 4 \pi f_\pi $, etc.

A finite  range of applicability may imply
not only the presence of new interactions, but may signal a radical
modification of the physical principles involved. A good example in
this respect concerns classical hydrodynamics. This is a very good theory at
scales larger than the interatomic distance, but when physics at smaller
distances is studied, a radical modification of the manner in which the
fluid is studied is required: quantum mechanics comes into play.
Though this comment is probably not relevant for most of the processes
to be studied in the forseeable future, it should be kept in mind when
thinking about the next great frontier, the gravitational interactions.

The procedure followed in all of the above examples in order to obtain
the effective lagrangian is straightforward: determine the particles involved
and the symmetries they are to obey, then construct the most
general set of local interactions containing the corresponding fields
constrained only by the said symmetries. (In practical
calculations one determines the most
important terms in this -- in principle infinite -- set of interactions
neglecting the rest.)
The lagrangian thus obtained has many dimensional parameters, some of these
can be associated with the masses and other scales of the low energy
physics. Other dimensional parameters, however, reflect the
scale of the high energy physics underlying the effective lagrangian.
It is because of this that current measurements can provide some
information about the scale of new physics.

Effective lagrangians can be used in perturbative calculations, just as
the ``usual'' lagrangians. As will be shown, the inevitable divergences
which arise can be dealt with in the same way. The reason why a
bad name has been sometimes associated with effective lagrangians is
not related to their inability to deal with these divergences (since
they can be treated in the same way as in the ``usual'' lagrangian case),
but their
alleged lack of predictability: these models contain an infinite
number of coefficients, so that an
infinite set of data points is apparently required in order to generate
accurate predictions from such a model. The way out of this apparent
deficiency
is based on the fact that the terms in the effective lagrangian
can be arranged
according to a certain hierarchy such that the lower order terms
generate the dominant contributions to any observable. Therefore,
to any order in this hierarchy, there is only a finite number of
operators which needs to be considered; one can also
estimate the corrections produced by
the terms which are neglected. Given an experimental result
one can truncate the effective lagrangian calculation whenever the
terms give contributions of the same order as the experimental
uncertainty.
Viewed in this light effective lagrangians are perfectly acceptable
theories which do produce non-trivial predictions;
moreover, they also provide the means of estimating the accuracy
of such predictions.

\REF\collins{J.C.  Collins, {\it Renormalization} (Cambridge U. Press,
Cambridge
             1984).}
\REF\bw{W. B\"uchmuller and D. Wyler, \npb268
        (1986) 621; see also W. B\"uchmuller \etal, \plb197
         (1987) 379.}
\REF\burges{C.J.C.  Burges and H.J. Schnitzer, \npb228 (1983) 464.}
\REF\keung{C.N. Leung \etal, \zphys31 (1986) 433.}
Understanding how  effective interactions are generated
is simple. Suppose the effective action
for a model is obtained; assume now, for example, that a dimensional
parameter is very
large (alternatively suppose we are interested in energies
much smaller than this scale). Expanding the effective action
in inverse powers of this
parameter generates an infinite set of local operators to be identified
with the terms in the effective lagrangian.  The full low energy
content of the model is then described by this object.
When the said
parameter becomes large, several fields will acquire large masses;
if the lagrangian with these fields removed is renormalizable
then the decoupling theorem ~\refmark{\collins} is applicable, and any
observable can
be expanded in inverse powers of the large dimensional parameter;
in particular all effects of the heavy physics disappear when the
dimensional parameter becomes infinite.
In this case the above
mentioned hierarchy in the effective lagrangian is determined by the
expansion in inverse powers of the large scale; for example, the
classification of the
effective operators of dimension six with \sm\ fields
was obtained in Refs.~\bw,\burges,\keung.

One can also consider the effect of having a large
dimensionless parameter in the original lagrangian. Then the effective
action can again be expanded in a series of local (effective) operators
which determines then the effective lagrangian describing the low
energy region of the model. When a dimensionless
coupling is large some fields might also become heavy (note
however that the {\it scale} of the corresponding
masses is unaltered). In this case the decoupling scenario is not
realized, and the hierarchy in the effective lagrangian is generated,
in many physically interesting cases, by a derivative
expansion.

The above two situations (to be denoted
the {\it decoupling} and {\it non-decoupling} cases respectively)
are quite different
qualitatively and quantitatively. In the
non-decoupling case the effects of a large (dimensionless) parameter
do not disappear (and some observables may even diverge) in the limit
when it becomes infinite. An
example is the effects of a (formally) infinitely massive
top quark in the \sm\ (with the \vev\ fixed).
This is not the case in the
decoupling scenario, for example, all observable
effects of an extra $Z'$ vanish as $ m_{ Z' } \rightarrow \infty $.

The reason behind these different behaviours can be easily seen
(qualitatively) by looking at the lagrangian. Setting a parameter
to infinity imposes a constraint: that the operator that this
parameter multiplies should vanish,\foot{{\ninerm
This discussion is purely classical, the quantum treatment of this
problem lies beyond the scope of this course and I refer the reader
to the literature~\refmark{\collins}.}}
else there will be associated with this term an infinite energy.
If the parameter corresponds to a mass then in the limit
the field it multiplies will vanish; if the parameter is
a mass term with the ``wrong'' sign (or a super-renormalizable
coupling) then the constraint forces the fields
multiplying the parameter to acquire a very large \vev, and this
implies that all fields getting a mass through this \vev\ should vanish.
In contrast, when a dimensionless parameter multiplying
an interaction becomes infinite, the resulting constraint imposes
a non-linear relationship among various fields and renders the low
energy theory non-renormalizable.

\REF\weinberg{S. Weinberg, {\it Physica} 96{\bf A} (1979) 327.}
\REF\georgi{H. Georgi, \npb361 (1991) 339; \ibid 363 (1991) 301.}
\REF\appelquist{T. Appelquist, \prd22 (1980) 200.}
\REF\bernard{C. Bernard, \prd23 (1981) 425.}
\REF\longhitano{A. Longhitano, \npb188 (1981) 118.}
\REF\appelquistwu{T. Appelquist and G.-H. Wu, \prd48 (1993) 3235.}
To illustrate these possibilities consider the large Higgs mass
limit of the \sm. The Higgs can be heavy either because
the \vev\ is big,
or because the scalar self interaction coupling $ \lambda
\rightarrow \infty $. In the first case
one must remember that the gauge boson and fermion masses also
increase without bound, so that the low energy theory contains
only photons and neutrinos. If
in contrast we let $ \lambda \rightarrow \infty $,
the Higgs mass again becomes infinite (though the gauge boson masses are
fixed), but the coupling between the
would-be Goldstone bosons also becomes infinite. To elucidate this situation
further consider the scalar doublet $ \Phi $ and construct the matrix
$ \Omega = ( \Phi , \tilde \Phi ) $, where $\tilde \Phi = i
\sigma_2 \Phi^* $, which satisfies $ \Omega^\dagger \Omega
= ( \Phi^\dagger \Phi ) {\bf1}$. I can then write
$$ \Omega = \sqrt{ \Phi^\dagger \Phi } \; U; \qquad
U^\dagger U = {\bf 1 } . \eqn\eq $$
The whole of the
\sm\ can be written in terms of $ \Omega $; for example
the scalar kinetic energy is $ \tr ( D \Omega )^\dagger \cdot
D \Omega $ where $ D_\mu \Omega= \partial_\mu \Omega
+ { i \over 2 } g W_\mu^a \sigma^a \Omega + { i \over 2 } g' B_\mu \Omega
\sigma_3 $\foot{{\ninerm This way of presenting things has many advantages,
for example, it is immediate that the scalar sector has an $ \su2_L
\times \su2_R $ invariance broken only by the coupling to the $ \ui $ gauge
field $B_\mu $.}}. The scalar potential is $$ V = \lambda
\left( \Phi^\dagger \Phi - \half
v^2 \right)^2 \eqn\eq $$ so that the constraint imposed
for large $ \lambda $ is $ \Omega = v U / \sqrt{2}  $ and the scalar sector
becomes, in the roughest approximation, a non-linear sigma model.
This is
the so-called non-linear or chiral realization
of the symmetry breaking in the
\sm~\refmark{\weinberg,\georgi,\pich,\appelquist,\bernard,\longhitano}.
The lowest order
effective operators in this scenario have been obtained by several
of the above references, I will use the conventions of Ref.~\appelquistwu.
In the following I will refer to the effective lagrangian constructed
from the gauge and fermion Standard-Model fields together with $U$,
as the {\it chiral} case. Note that there are no physical scalar
excitations in this scenario.

\REF\yao{H. Steeger \etal, \prl\ {\bf59} (1987) 385.
         G.-L. Lin \etal, \prd44 (1991) 2139; University of Michigan
         report UM-TH-93-05 (unpublished).}
There are, of course, various other limits of interest.
One can imagine letting a
fermion mass increase by making a Yukawa coupling large~\refmark{\yao}. In this
case,
just as in the previous
situation, several interaction strength become large also. The result
is a kind of non-linear realization of the symmetry involving the
fermions as well as the scalars. It might seem that these
complications disappear, at least in the \sm, as soon as we
imagine letting the Yukawa couplings of both elements of a doublet
become large, but this is not so. First, since we are letting a dimensionless
parameter become large, large fermion-scalar
couplings are generated. Second, the theory without a doublet is not
consistent, but in order to explain why I need to mention what an anomaly is.

\REF\tjg{S. Treiman \etal, {\it Lectures on Current Algebra and Its
Applications} (Princeton Univ. Press, Princeton, N.J. 1972).}
\REF\jackiwgross{ R. Jackiw, and D. Gross, \prd6 (1972) 477.}
The symmetries of a model are associated, via Noether's
theorem~\refmark{\itzykson,\tjg}, to
a series of conserved quantities. It is then usually assumed that the
corresponding symmetries are preserved after quantization; but this
may not be the case. There are many situations (usually
involving fermions) in which quantum corrections spoil certain
symmetries.
By this I mean that given an operator $
j_\mu $ whose classical counterpart is conserved, it does not follow that
the matrix elements of the operator $ \partial \cdot j $ will vanish.
They will of course
be zero to $ O ( \hbar ) $ since this is the classical
result, but in certain
situations the $ O ( \hbar ) $ contributions do not vanish~\refmark{\tjg}. In
this
case the current $ j_\mu $
and associated symmetry are labelled anomalous.
If the anomalous current is also
a gauge current then the theory
is inconsistent~\refmark{\jackiwgross}, at least within perturbation theory.

\REF\farhidhoker{E. D'Hoker and E. Farhi, \npb248 (1984) 59,
                 \ibid 77. See also T. Sterling and M. Veltman
                 \npb189 (1981) 557.}
\REF\wess{ J. Wess and B. Zumino, \plb37 (1971) 95.
           E. Witten, \npb223 (1983) 422.}
Returning to the previous discussion we imagine letting the (common)
Yukawa coupling
$y$ of a \sm\ fermion doublet to be very large~\refmark{\farhidhoker}.
The doublet masses will then be very large, but the \sm\ without a doublet
has an anomalous gauge symmetry and is therefore inconsistent.
What explicit calculations show is that upon letting
$ y \rightarrow \infty $, a tower of CP violating scalar interactions
is generated which build up to the so-called
Wess-Zumino lagrangian~\refmark{\wess}. This function of the
scalar field is such that it generates the same anomaly as the
apparently decoupled doublet. So the contributions from the light fermions
plus the one produced by the WZ lagrangian render the
theory non-anomalous. There are also other remnants of the heavy
doublet. For example the contributions to the slope of the vacuum
polarization of the vector bosons goes to a constant as $y$
becomes infinite, and this slope (essentially the
so-called $S$ parameter) is measurable.

\section{{\it Effective vs. renormalizable lagrangians.}}

Many very successful field theories depend on a small number
of unknown parameters. For example QED for photons and electrons has two
parameters: the electron's mass and charge. The standard electroweak
model has (ignoring topological terms)
three gauge constants, two parameters in the scalar potential,
nine masses and four parameters in the KM matrix: a total of 19
unknown parameters. The reason these models can get away with a finite
number of parameters is tied to their renormalizability, which
I describe next.

A theory is renormalizable if, in units where $ \hbar = c = 1 $,
all terms in the lagrangian have mass dimensions $ \le 4 $ and if
all propagators vanish at large energies. For theories with particles
with spin $ \ge 1 $ this last constraint requires some sort of
gauge invariance (else the propagators either don't exist, as for
massless particles, or else they go to an energy independent tensor
at large energies, as is the case for massive particles).

The only difference between effective and renormalizable lagrangians
lies in the allowed dimensions of the allowed terms: in
effective theories there is no restriction whatsoever.
Gauge invariance is still imposed for particles of spin $ \ge 1 $
in effective theories: it is required by  unitarity. It is a well
known fact that theories with particles with spin $ \ge 1 $ posses
states of negative probability~\refmark{\itzykson} and that the only known
method of
neutralizing the noxious effects of these states is based on a
gauge principle (Gupta-Bleuler formalism, Fadeev-Popov technique,
etc.)

If all allowed operators in an effective theory
were equally important such a model would
be useless (even writing down the lagrangian would be
an impossible task). Fortunately the operators can be ordered so
that, within the range of applicability of the model,
higher order terms will generate small corrections to the results
obtained using lower order terms. For example,
if the heavy physics
decouples, then all operators will be
suppressed by the appropriate power of the heavy mass $ \Lambda $ and
the effective lagrangian can be organized as a power series in
$ 1 / \Lambda $; this is also true for all observables.
Thus when considering processes at energies
$ \ll \Lambda $, the higher dimensional operators will generate,
in general, small corrections.

In the non-decoupling case the heavy physics is heavy due to the presence
of a large dimensionless coupling . The resulting
expansion is different: the decoupling theorem does not
apply and, therefore, we can expect contributions to the effective lagrangians
which do not vanish as the heavy mass increases. What can be done is then
to present an expansion in powers of the external momenta or, equivalently,
a derivative expansion~\refmark{\weinberg}.

Given a lagrangian (effective or renormalizable) we have a set of
undetermined parameters. When calculating the value of any observable
the result will be a function of these parameters. Thus one can
choose a (sufficiently large) number of observables and invert
these relations; all parameters of the theory will then be expressed in
terms of this set of observables, the ``input data''. The theory can
now be used to predict the values of other observables in terms of
the input data. As a technical detail I must point out that
the calculation of observables cannot be done exactly and that
some approximation technique is used.

A problem with the above program is that some calculations will be
infinite. This is usually associated with integrals diverging in
the UV limits of integration. The simplest
example corresponds to the energy of the vacuum of a free bosonic field:
the field corresponds to an infinite set of harmonic
oscillators~\refmark{\itzykson}
and so the vacuum energy diverges.
To be able to handle such diverging
quantities the model is regularized (for a clear discussion see
Ref.~\collins). This means that the model under
consideration is understood as the limit of a family of models such
that for each member of the family the results are finite, while the
divergences are recovered in the limit. Then, provided all relevant
symmetries are obeyed by all members of the family, one can work with
the lagrangians which give finite results, and take the limit at the
end of any calculation (after all observables are expressed in terms of
the input data).

For example one can require that the Fourier expansion of any field
is cut off at a scale $ \Lambdauv $, and take the limit $ \Lambdauv
\rightarrow \infty $ at the end of the calculation. This regularization
prescription is problematic because one cannot in general preserve
gauge invariance and Lorentz invariance.
A more attractive possibility is to modify the kinetic term in the lagrangian.
For example, in the case of a neutral scalar field $ \phi $ the
modification is
$$ \left( \partial \phi \right)^2
\rightarrow \left( \partial_\mu \phi \right)
\left[ 1 + ( \square /\Lambdauv^2 )^n \right]
\left( \partial^\mu \phi \right) \eqn\eq $$ where $n$ is a sufficiently large
integer. The problem with this method is that the propagator
acquires a set of extraneous poles whose residues have the
``wrong'' sign; these represent states of negative norm. The scale of
these unwanted states is $ \Lambdauv $ so one has to prove that these
poles leave no remnants when $ \Lambdauv \rightarrow
\infty $.

A better method is to
define the lagrangian in an arbitrary number of dimensions $n$,
perform all calculations, and then take the limit $ n \rightarrow 4 $.
This scheme is called dimensional regularization.
In the first regularization techniques divergences are substituted
by powers or logarithms of $ \Lambdauv $, in the last scheme they appear
as terms proportional to $ 1 / ( n - 4 ) $ to some power.

So what is usually done is to first regularize the theory, then re-write
all expressions in terms of the chosen set of observables (the input data).
Though I will not prove it here, the expressions for all
observables will then be finite in the limit where the regulator disappears
(\ie\ when the cutoff is set to infinity, the dimension of space-time is
taken to four, etc.)~\refmark{\collins}.
For renormalizable lagrangians the substitution
of lagrangian parameters for observables requires a finite set of
operations, for the effective lagrangians the number of operations is
(in principle) infinite.

For example, consider the graph below

\setbox2=\vbox to 2 truein{\epsfxsize=5 truein\epsfbox[0 0 612 792]{figm0.ps}}
\centerline{\box2}
\vskip -50pt

\noindent representing a correction to the
electron vacuum polarization in QED. Among other effects this graph
shifts the pole in the electron propagator
away from its original value $m_0$: now the pole is at
$ m_0 + \alpha_0 m_0 \times $(divergent quantity)$ = Z_m m_0 $
\foot{{\ninerm The correction is also proportional to $m_0 $ due to
chiral symmetry
which prohibits the generation of a mass to all
orders in perturbation theory. The quantity $ Z_m $ diverges as
the regulators are removed, \ie, as $ \Lambda_{ UV } \rightarrow \infty $,
$ n \rightarrow 4 $, etc.}} (where $ \alpha_0 $ is a parameter in the
lagrangian
describing the coupling of the electrons to the photons). It
is then $ m = m_0 Z_m $ that is measured in the laboratory as
$ 9.1093897 \times 10^{ - 31 } \kg $ while $ m_0 $ by itself has no
direct physical significance. Similarly $ \alpha = Z_\alpha \alpha_0 $
and $ \alpha = 1/ 137.036 $. When QED is dimensionally regularized,
for example, the $Z$ factors are expressed as Laurent series
in $ n - 4 $; in fact they
can be chosen as a power series in $ 1/ ( n - 4 ) $ with coefficients
depending on $ \alpha $ but independent of $m$. According to what I
stated previously, when any physical prediction is written in terms of
$m$ and $ \alpha $ (as opposed to $ m_0 $ and $ \alpha_0 $) all the results
are regulator independent and finite. The only point I glossed over
pertains the possibility of rescaling the fields but, when $S$ matrix
elements are calculated, this is irrelevant.

\section{{\it Symmetries and construction of the effective lagrangians.}}

The ingredients required for the construction of an effective lagrangian
are the relevant fields and the symmetries they are to obey.
The requirement that the effective lagrangian be invariant under
certain symmetries (and no others) can have important consequences. For
example,
a low energy description of the strong interactions should satisfy
$ \ccal, \ \pcal $ and $ \tcal $ independently. But when the
simplest lagrangian
for the lowest meson octet is constructed ,it is found to have
two different operations associated with $ \pcal $.
This suggests that the simplest effective
lagrangian is incomplete since it lacks the terms
which break this group to the usual $ \ZZ_2 $.
This is in fact the case; the required terms are generated by the
Wess-Zumino lagrangian~\refmark{\wess} which describes a
wealth of phenomena previously unaccounted for,
such as the decay of neutral pions into two photons,
Thus, after one has constructed the most
general set of operators with the chosen fields obeying the required
symmetries, then one must verify that there are no
spurious symmetries; if present this suggest possible terms have been omitted.

\REF\burgessandlondon{C.P. Burgess and D. London, McGill University report
                      MCGILL-92-04 (unpublished);
                      (Bulletin Board: hep-ph@xxx.lanl.gov - 9203215).}
\REF\wudkarev{J. Wudka, UC Riverside report UCRHEP-T121}
A prominent role among symmetries is occupied by gauge symmetries. This is
so because, as mentioned above, they insure unitarity.
It is however true that an arbitrary lagrangian can be understood as
the unitary gauge limit of a gauge invariant lagrangian.
This is shown by explicitly
constructing the said  gauge invariant
lagrangian which contains in addition to the original fields
a set of auxiliary fields~\refmark{\burgessandlondon}; such models are
in general non-renormalizable.
Thus it appears that gauge invariance is of no importance
(since any theory can be thought as a gauge theory, albeit in the unitary
gauge). There are, however, problems with this statement~\refmark{\wudkarev}.
First, the procedure by which
a model is rendered gauge invariant does not fix the group
(so that, for example, if we are given the
\sm\ lagrangian in the unitary gauge
we can turn it into an $ \su2 \times \ui $ {\it or} a $ \ui^4 $ gauge
theory). Second, the gauge transformation properties of the matter fields
are not fixed in this procedure. When the (experimentally supported)
transformation properties of the matter fields and the values of the
gauge couplings are imposed the (now gauge invariant)
model acquires content, but then the
requirement that it is gauge invariant is a no longer trivial.

Once gauge invariance is imposed on an effective lagrangian the number
of allowed operators is severely limited.
For example~\refmark{\wudkarev}, the coupling of the $W$ bosons
to the $Z$ and photon is described, in the decoupling case
to order $ 1 / \Lambda^2 $, by four parameters instead of the
seven allowed by imposing only Lorentz invariance. Again to this order,
the couplings among four vector fields are described by two parameters, etc.

\section{{\it Equations of motion.}}

\REF\arztiii{C. Arzt, Univ. of Michigan report  UM-TH-92-28 (unpublished)
             (Bulletin Board: hep-ph@xxx.lanl.gov - 9304230).}
It is often found that two effective operators differ only by terms that
vanish when the classical equations of motion are assumed.
In this case the S matrix cannot distinguish between them;
this is
trivial in the case of tree level diagrams involving one insertion of
these operators, but is also true for loop graphs as I will now
prove;
I will restrict myself to the simple case of a scalar theory
(for a complete discussion see Refs.~\georgi,\arztiii).
I will denote the fields by $ \phi $ and the classical action by
$ S ( \phi ) $.

Suppose that we have two operators $ \ocal $
and $ \ocal' $ such that $$ \ocal ' = \ocal + \int d^4 x\; \acal
{ \delta S \over \delta \phi } \eqn\eq $$ for some local quantity
$ \acal $ depending on the $ \chi $. The effective action is
$$ S \lowti{ eff } = S + \int d^4 x \; \left( { \alpha
\over \Lambda^2 }  \ocal +
{ \alpha' \over \Lambda^2 }
\ocal' \right) + \cdots ; \eqn\eq $$  the dots indicate higher
dimensional operators. Let $ S ' = S + \int d^4 x \left[
( \alpha + \alpha' ) / \Lambda^2 \right] \ocal $,
then $$ \eqalign{ S \lowti{eff }
&= S' ( \phi ) + {\alpha' \over \Lambda^2 } \int d^4 x\;
\acal { \delta S \over \delta \phi } + \cdots \cr
&= S' ( \phi ) + { \alpha' \over \Lambda^2 } \int d^4 x\;
\acal { \delta S' \over \delta \phi } + \cdots \cr
&= S' ( \phi + \alpha' \acal / \Lambda^2 ) + \cdots \cr } \eqn\eq $$
Thus, to the order we are working, the effects of $ \ocal' $ are
to replace $ \alpha \rightarrow \alpha + \alpha' $ and $ \phi \rightarrow
\phi + \alpha' \acal / \Lambda^2 = \Phi$.
Using $$ \phi = \Phi - { \alpha' \over \Lambda^2 }
\acal ( \Phi ) + \cdots \eqn\eq $$ the effects of
replacing $ \phi \rightarrow \Phi $ appear only in external legs
as in the following figure

\setbox2=\vbox to 2 truein{\epsfxsize=5 truein\epsfbox[0 0 612 792]{figm1.ps}}
\centerline{\box2}
\vskip -50pt

\noindent where light lines denote $ \phi $ propagators and heavy lines
denote $ \Phi $ propagators. Dots indicate external legs.

The S matrix is obtained by amputating the external legs and
putting the resulting expression on the mass shell; graphically

\setbox2=\vbox to 3 truein{\epsfysize=5 truein\epsfbox[0 0 612 792]{figm2.ps}}
\centerline{\box2}
\vskip -50 pt

But the terms in a Green's function
with two or more $ \Phi $ legs emanating from the
same external point are regular when that external point is
put on mass shell; that is

\setbox2=\vbox to 2.5 truein{\epsfxsize=4.25 truein\epsfbox[0 0 612
792]{figm3.ps}}
\centerline{\box2}
\vskip -50 pt

\noindent These regular terms, when multiplied by the corresponding
inverse propagator,
will vanish when the mass shell condition is imposed.
It follows that only the terms linear in $ \Phi $
in the expansion of $ \phi $ will contribute
to the S matrix. Thus, if near the mass shell $ \phi = z \Phi + \cdots $
for some constant $z$ (the dots indicate terms with more
$ \Phi $ fields or terms which vanish on mass shell) then
the effect of the operator $ \ocal' $ is then the replacement $ \alpha
\rightarrow \alpha+ \alpha' $ and the appearance of the wave
function renormalization factor $z$. This last quantity is unobservable:
the precise same factor appears in the propagator
(near the mass shell), so when the lines are amputated the $z$
factors cancel out. I can then conclude that we can take $ \ocal' $
into account by the simple replacement
$ \alpha \rightarrow \alpha + \alpha' $.

\chapter{Tree level applications.}


\REF\burgesschnitzer{C.J.C.  Burges and H.J. Schnitzer, \plb134 (1984) 329.}
\REF\doncheski{K.A. Doncheski,\zphys52 (1991) 527.}
\REF\wudkahera{J. Wudka, Univ. of California Riverside report T113 and
               \prd (to appear).}
The simplest application of effective lagrangians appear in
tree level processes\foot{{\ninerm It
must be emphasized that this means tree level processes in the
effective lagrangian, the effective operators summarize the low
energy limit of loop graphs involving heavy particles. }} which
I will describe with an example.
Consider the process $ p e_R \rightarrow \nu_L X $
which can
be generated at HERA~\refmark{\burgesschnitzer, \doncheski, \wudkahera}.
In the \sm\ this is a very suppressed
reaction: it requires a helicity flip and will therefore
be proportional to the electron mass divided by the CM energy.
On the other hand there are several dimension six operators
that can contribute to this process whose effects might be important.
I will assume that the heavy physics is weakly coupled.

The relevant four-fermi operators are
$$ \ocal_{ \ell q } = ( \bar \ell e ) \epsilon ( \bar q u ), \quad
\ocal_{ q d e } = ( \bar \ell e ) (\bar d q ), \quad
\ocal_{ \ell q }' = ( \bar \ell u ) \epsilon ( \bar q e ) ,\eqn \theops $$
where I have adopted the notation and conventions of B\"ucmuller
and Wyler~\refmark{\bw}: $
\ell $ and $q$ denote the left-handed lepton and quark doublets respectively, $
e, \ u $ and $d$ denote the right-handed electron, up and down quark fields,
and $ \epsilon = i \sigma_2 $. The graphs for the process at hand
containing these operators are simply

\setbox2=\vbox to 1.5 truein{\epsfysize=4 truein\epsfbox[0 0 612
792]{figm4.ps}}
\centerline{\box2}
\vskip -50pt

The remaining operators containing \sm\ fields which contribute to the process
at hand are $$
\ocal_{ D e } = ( \bar \ell D_\mu e ) ( D^\mu \phi ), \quad
\ocal_{ \bar D e } = ( \overline{ D_\mu \ell } e ) ( D^\mu \phi ), \quad
\ocal_{ e W } = g \bar \ell \sigma^{ \mu \nu } \sigma^I e \phi W^I_{ \mu \nu }
\quad \eqn\moreops $$ where $ \phi $ denotes the scalar doublet, $ D_\mu $ the
covariant derivative, $ \sigma^I $ the Pauli matrices, and $ W^I_{ \mu \nu
}$ the $ \su2 _L $ gauge field strength tensor with $g$ the corresponding
coupling constant. These operators contribute via $t$-channel $W$ exchange
via the graphs

\setbox2=\vbox to 1.5 truein{\epsfxsize=4 truein\epsfbox[0 0 612
792]{figm5.ps}}
\centerline{\box2}
\vskip -50 pt

The lagrangian is therefore $$ \leff = \lcal\lowti{St. Model} + \inv{
\Lambda^2 } \sum_i \alpha_i \ocal_i \eqn\leftl $$ (the sum over $i$ runs over
the above six terms). I
have kept only the operators of lowest dimension contributing at tree level, as
they give the leading contributions.

There are some experimental restrictions
on the coefficients $ \alpha_i $~\refmark{\bw}: from
$K$ and $ \pi $ decays it is known that $$ \alpha_{ q d e } \simeq 0
\qquad \hbox{and} \qquad \alpha'_{ \ell q } \simeq 2 \alpha_{ \ell q }
\eqn\constr $$ which I will adopt;
these conditions are assumed to be the result of some (unknown) constraint
stemming from the underlying physics. With these restrictions the
four-fermion interactions correspond to a tensor exchange,
$$ \ocal_{ \ell q } + 2 \ocal_{ \ell q }' \propto
\left( \bar \nu_L \sigma_{ \mu \nu } e_R \right)
\left( \bar d_L \sigma^{ \mu \nu } u_L \right) . \eqn\eq $$.

With \leftl\ we can calculate the cross section for the process at hand. Note
that the two types of operators \theops\ and \moreops\ will not interfere due
to helicity conservation at the quark vertex (provided quark masses
are ignored). The result is
$$ \eqalign{ { d \sigma_L \over dx dy } =& { | \alpha_{ \ell q } |^2
s \over 32 \pi \Lambda
^4 } x \left[ (2 - 3 y )^2 U + ( 2 - y )^2 \bar D \right] \cr & + \inv{ 32 \pi
}
\left| { g^2 v \over \sqrt{8} } V_{ d u } \up{ K M } { c \over \Lambda^2 }
\right|^2 \left[ { x^2 y ( 1 - y )
\over ( x y + m_w^2/s )^2 } ( U +
\bar D ) \right] \cr} \eqn\sigl $$ where $x$ and $y$ are the usual scaling
variables, $ v = \sqrt{2} \langle \phi \rangle \simeq 256 \gev $, $U$ and $
\bar D $ are the ($x$ and $y$-dependent) quark distribution functions; I
also
defined $c =  \alpha_{ De } - \alpha_{ \bar D e } + 8
\alpha_{ e W } / g $. I will show later that when the underlying physics
is weakly coupled $ c \sim 1/ ( 16 \pi^2 ) $ and $ \alpha_{ \ell q } \sim 1
$.

Using \sigl\ we can estimate the sensitivity of HERA to the scale $ \Lambda $:
the above cross section will generate 15 events per year at HERA provided $
\Lambda \lesim 315 \gev $.

This result will be weakened for polarizations below 92\%; in this
case the statistical significance of the right-handed electron signal
must be considered. Suppose that we have a beam of polarization $ \pcal
$, where $ \pcal = 1 $ implies pure right-handed electrons. Let $
\sigma_{ SM } $ be the \sm\ contribution to $ e_L p \rightarrow \nu_L
X $ via $t$-channel $W$ exchange; a simple calculation shows that
\foot{{\ninerm There is a large class of operators which also modify
the couplings of the $W$ to the left-handed weak currents, as well as
shifting the $W$ mass from it's standard model value. I have
not included these contributions in $ \sigma_{ SM } $ since they
represent but small corrections.}}
$$ \sigma_{ SM } = { g^4 \over 8 \pi s } \int_0^1
dx dy { x U + x ( 1 - y ) ^2 \bar D \over \left( xy + m_W^2/s \right)^2 }
\simeq5.67 \times 10^{ - 7 } \gev^{ - 2 } , \quad ( \sqrt{s} = 292 \gev )
\eqn\eq $$ The number of signal events is $ N \lowti{ signal } = \pcal \sigma_L
\lBB $, where $ \lBB $ is the luminosity ($ \lBB = 4 \times 10^9 \gev^2 /$year
for HERA); the number of background events is $ N\lowti{bckgnd} = ( 1- \pcal )
\sigma_{ SM
} \lBB $; the condition for the signal to be statistically significant is $$ N
\lowti{ signal} > \sqrt{ N\lowti{bckgnd}
 + N \lowti{ signal } } , \eqn\statsign $$ which
determines the sensitivity to $ \Lambda $ given $ \pcal $. This condition
generates the following graph

\setbox2=\vbox to 3.1 truein{\epsfxsize=5 truein\epsfbox[0 0 612
792]{figm6.ps}}
\centerline{\box2}
\vskip -50 pt

\noindent
where I assumed $
\left| \alpha_{ \ell q } \right| = 0.44 $. This plot
gives the maximum sensitivity to $ \Lambda $ for a given  polarization.
For $ \pcal \le 0.9 $ the curve is
generated by \statsign, for $ \pcal > 0.9 $ the
curve corresponds to 15 events
per year at HERA. For realistic values of $\pcal$ the
process is no longer rate dominated.

\section{{\it S,T, U.}}

\REF\peskin{M. Peskin and T. Takeuchi, \prl65 (1990) 964.
            G. Altarelli and R. Barbieri, \plb253 (1991) 161.
            B. Lynn \etal, in {\it Physics at LEP}, CERN
            Yellow report 86-02.}
Some of the best measured quantities which are sensitive to new physics
are the oblique parameters
$S$, $T$ and $U$~\refmark{\peskin} obtained from the vacuum polarization
tensors for the $W$ and $Z$ bosons.
In terms of the $\su2 \times \ui $ eigenstates
$$ \eqalign{
S &= - 16 \pi \Pi'_{ 3 B } ( 0 ) , \cr
T &= { 4 \pi \over ( \sw \mw )^2 } \left[ \Pi_{ 1 1 } ( 0 ) -
\Pi_{ 3 3 } ( 0 ) \right] ; \cr
U &= + 16 \pi \left[ \Pi'_{ 1 1 } ( 0 ) -
\Pi'_{ 3 3 } ( 0 ) \right] ; \cr } \eqn\eq $$
where the indices $ 1 $ and $3$ refer to $ \su2 $, the subindex $B$ refers to
$  \ui $. The functions $ \Pi $ are the transverse part of the vacuum
polarization tensors, they are functions of $ p^2 $;
the prime indicates a derivative with respect to $ p^2 $.

The \sm\ contributions have been studied
extensively in the literature; concerning the possible contributions
form new physics I will use an effective lagrangian parametrization
and consider the operators which have two gauge bosons, no fermions
and any number of scalars.
For the decoupling case these are~\refmark{\bw}
$$ \eqalign{
\ocal_{\phi W } &= \half \left( \phi^\dagger \phi \right)
\left( W_{ \mu \nu }^I \right)^2 ; \qquad
\ocal_{\phi \tilde W } = \half \left( \phi^\dagger \phi \right)
\left( W_{ \mu \nu }^I \tilde W_{ \mu \nu }^I \right) ; \cr
\ocal_{\phi B} &= \half \left( \phi^\dagger \phi \right)
\left( B{ \mu \nu } \right)^2 ; \qquad
\ocal_{\phi \tilde B} = \half \left( \phi^\dagger \phi \right)
\left( B{ \mu \nu } \tilde B{ \mu \nu } \right) ; \cr
\ocal_{ W B } &= \left( \phi^\dagger \sigma_I \phi \right)
W_{ \mu \nu }^I B^{ \mu \nu } ; \qquad
\ocal_{ \tilde W B } = \left( \phi^\dagger \sigma_I \phi \right)
\tilde W_{ \mu \nu }^I B^{ \mu \nu } ; \cr
\ocal_\phi \up 1 &= \left( \phi^\dagger \phi \right)
\left| D_\mu \phi \right|^2 ; \qquad
\ocal_\phi \up 3 = \left| \phi^\dagger D_\mu \phi \right|^2 ; \cr }
\eqn\decstuops $$
so that the effective lagrangian is $$ \leff = \lcal\lowti{St. Model}
+ \sum_i { \alpha_i \over \Lambda^2 } \ocal_i \eqn\eq $$
where the sum over $i$ runs over the above operators.
I then find  $$
S = { 2 v^2 \over \pi \Lambda^2 } \left( { 16 \pi^2 \alpha_{ W B }
\over g g' } \right) ; \quad
T = - { 16 \pi v^2 \over \sw^2 \Lambda^2 } \alpha_\phi \up 3; \quad
U = 0 . \eqn\eq $$

For the chiral case the relevant effective lagrangian
is~\refmark{\appelquistwu}
$$ \eqalign{
\leff =& { \beta_1 g^2 v^2 \over 4 } \left( \tr \left\{ \sigma_3
U^\dagger D_\mu U \right\} \right)^2 + { \alpha_1 g g' \over 2 }
B^{ \mu \nu } W^I_{ \mu \nu }
\tr \left\{ U^\dagger \sigma_3  U \sigma_I \right\} \cr &  + {
\alpha_8 g^2 \over 4 } \left( W_{ \mu \nu } ^I \tr
\left\{ U^\dagger \sigma_3  U \sigma_I \right\} \right)^2  ; \cr}
\eqn\eq $$ whence $$
S = - \inv\pi \left( 16 \pi^2 \alpha_1 \right) ; \quad
T = \inv{ 2 \pi \sw^2 } \left( 16 \pi^2 \beta_1 \right) ; \quad
U = - \inv\pi \left( 16 \pi^2 \alpha_8 \right) . \eqn\eq $$

I will prove later that the natural size for the coefficients
$ \alpha_{ 1 , 8 } $ is $ 1 / ( 4 \pi )^2 $; the natural magnitude of
$ \beta _1 $ is $ \lesim 1 $ though experimental constraints
requires $ \beta_1 \lesim 1/ ( 16 \pi^2 ) $.
Comparing these two sets of expressions
I can conclude that if the operators
of dimension six in the effective lagrangian dominate (for the
decoupling case) then I expect $ U \simeq 0 $; in contrast, for the
chiral, case $ U \sim O ( 0.1 ) $.

The operators in \decstuops\ are generated via loops
in the heavy theory, except $ \ocal_\phi \up{ 1 ,3 } $. Moreover
each $W$ will also be accompanied
by a $g$ and each $B$ by a $g'$. Then $ \alpha_{ W B } \sim g g'
/ 16 \pi^2 $ and $ \alpha_\phi \up3 \sim 1 $.
Thus I expect $ | S | \sim 2 v^2 / \pi \Lambda^2 $ and $ |T| \sim
16 \pi v^2 / ( \sw^2 \Lambda^2 ) $. For $ \Lambda \sim 1 \tev $
I get $ |S| \sim 0.15 $; using $ | T | \lesim 1 $ I get
$ \Lambda \gesim 10 \tev $

But there will be, in the decoupling case,
contributions to $S, \ T$ and $U$ form
tree-level-generated dimension eight operators. Then,
for example,
$$ S \sim\left[
\underbrace{ \left( { v^2  \over 16 \pi^2 \Lambda^2 }
\right)}_{\hbox{dim=6, 1  loop}} + \underbrace{
\left( { v^4 \over \Lambda^4 } \right) }_{ \hbox{
dim=8, tree level} }\right] \times 100 \eqn\eq $$
so that the dimension eight operators dominate if
$\Lambda < 4 \pi v \sim 3 \tev $.

For the parameter $U$ there are then two cases cases

$$ U \simeq \cases{ 0 & for $ \Lambda > 4 \pi v $ \cr
                   0.1 & for $ \Lambda < 4 \pi v $ \cr } \eqn\eq $$
The second cases comprising the both the chiral case and the
decoupling case when the dimension-eight operators are important.
The $U$  parameter would be then very useful in distinguishing among these
possibilities. As of now the measurements are to $ O ( 1 ) $ but
an improvement by a factor $ \sim 10 $ (and into the interesting region)
are to be expected from Fermilab and LEP in the near future.

\chapter{Propetries of the effective lagrangians}

\section{{\it Magnitude of the coefficients.}}

When considering
an effective operator it is important in quantitative estimates
to be able to predict the order of magnitude of its coefficients.
This allows for the determination of the dominant contributions
to any given observable as well as its sensitivity to
new physics effects. There are two cases of interest depending
on whether the underlying physics decouples or not.

\subsection{{\it Loops vs. tree level contributions.}}

 For the
case where the underlying theory is weakly coupled this implies
determining whether the operator in question is generated at
tree level by the heavy dynamics or whether it is loop generated;
in the second case, there will be a
suppression of $ \sim 1 /1 6 \pi^2 $ in the coefficient.
In addition there will be other suppression factors
due to the presence of weak ($ \lesim 1 $) coupling constants.

I will now describe how to determine whether an operator is generated
at tree level or at one loop. I will
assume that the theory underlying the \sm\ is a gauge theory;
the corresponding gauge indices are $a,b,$ etc. The gauge group,
of course, contains the \sm\ gauge group, whose indices will be denoted
$ I , \ J $, etc.

Consider for example
$$ \ocal_W = \epsilon_{ I J K } W_{ \mu \nu }^I
W_{ \nu \lambda }^J W_{ \lambda \mu }^K . \eqn\eq $$ If this operator
is generated at tree level there must be a graph of the form

\setbox2=\vbox to 2 truein{\epsfysize=5 truein\epsfbox[0 0 612 792]{figm7.ps}}
\centerline{\box2}
\vskip -50pt

\noindent This is so because gauge invariance in the light
degrees of freedom requires the full non-Abelian field tensor to appear
in $ \ocal_W $ which therefore contains a term with six external
light legs.  The only tree-level graph which generates such a term in a
Yang-Mills theory is then the one depicted above.

Note however, that $ f_{ b I J } = 0 $ when $b$ is the index of a heavy
gauge boson. This is true because the light vector bosons are the
gauge bosons of a bona-fide gauge theory, hence the commutator of two
generators with light indices will give a linear combination of generators
all with light indices. Hence the structure constants with two
light and one heavy indices vanish.
It follows that $ \ocal_W $ can only be generated by loop graphs
of the type.

\setbox2=\vbox to 2 truein{\epsfysize=3.5 truein\epsfbox[0 0 612
792]{figm8.ps}}
\centerline{\box2}
\vskip -50pt

As another example consider the four fermion operator
$$ \ocal= \left( \bar \psi_1 \gamma_\mu \psi_2 \right)
          \left( \bar \psi_3 \gamma^\mu \psi_4 \right) \eqn\eq $$
which can be generated by a heavy gauge boson exchange as in the
following graph

\setbox2=\vbox to 2 truein{\epsfysize=3.5 truein\epsfbox[0 0 612
792]{figm9.ps}}
\centerline{\box2}
\vskip -50pt

\REF\arztii{C.Arzt \etal, in preparation.}
\REF\wudkarev{J. Wudka, UC Riverside report UCRHEP-T121}
All operators appearing in Ref.~\bw\ can be analyzed in this way. I refer the
reader to
Refs.~\arztii,\wudkarev\ for a detailed discussion.

\subsection{{\it Strongly coupled case}}

\REF\georgibook{H. Georgi, {\it Weak Interactions and Modern
                Particle Theory} (Benjamin -- Cummings, Menlo Park, CA, 1984).}
\REF\georgimanohar{H. Georgi and A. Manohar, \npb234 (1984) 189.}
In order to determine the order of magnitude of the coefficients
of the effective operators
it is natural to
require that the radiative corrections
to the coefficient of an operator be at most as large as its tree level
value.

Consider a theory with scalar fields $ \phi $ and
fermionic fields $ \psi $  and gauge bosons $W$. Then the
relevant vertices have the symbolic form
$$ \Lambda^4 \lambda ( \phi / \lb )^A (\psi / \lf^{3/2}
)^B (p / \Lambda )^C ( g W / \Lambda )^D ,
\eqn\strongvertex $$
where $p$ represents a derivative, $  \Lambda $ is a UV cutoff,
$ \lambda $ is a coupling constant, and the other scales,
$ \Lambda_{ \phi , \psi } $, are to be
determined. Since $ \Lambda $ is associated with the momentum
scale I divide
$p$ (a generic momentum) by this scale; since gauge fields appear only in
covariant derivatives, they are divided by the same scale. The
quantities $A, \ B,  \ C $ and $D$ are assumed to be integers.
Since the $W$ fields appear always inside a covariant derivative
it is sufficient to consider vertices with $ D =  0 $.

Now consider a graph with $V$ vertices which generates an
$L$ loop correction to \strongvertex. This contribution
will be of
the same order provided (I replace all loop moment by
$ \Lambda $ since we are interested only in an order
of magnitude estimate)  $$ 1 \sim (
\Lambda^4 \lambda )^{ V -1 } \lb^{ A - \sum A_i } \lf^{3( B- \sum B_i )/2 }
\Lambda^{ C - \sum C_i }
\Lambda^{ \sum C_i - C + 4 L - 2 \ibos -\ifer }
( 4 \pi )^{ - 2 L } , \eqn\inter $$ where
$i$ labels the vertices in the graph and $ \ifer $ ($ \ibos $)
is the number of internal fermion (boson)
propagators. Using the relations $ \sum A_i = A + 2
\ibos $ and  $ \sum B_i = B + 2 \ifer $ \inter\ becomes
$$ ( 16 \pi^2 \lambda )^{ - L }
\left( { \lambda \Lambda^2 \over \lb^2 } \right)^\ibos \left( { \lambda
\Lambda^3 \over \lb^3 } \right)^\ifer \sim 1 ; \eqn\eq $$
this requires $$ \lambda \sim
\inv{ 16 \pi^2 } ; \qquad \lb \sim \inv{ 4 \pi } \Lambda ;
\qquad \lf \sim \inv
{ ( 4 \pi )^{ 2/3 } } \Lambda . \eqn\eq $$
Substituting back into \strongvertex, and using the fact that gauge bosons
appear always in a covariant derivative denoted by $ \dcal $, yields
$$ {\Lambda^4 \over ( 4 \pi )^{ 2 - A - B } }
\left( { \phi \over \Lambda } \right)^A \;
\left( { \psi \over \Lambda^{ 3/2 } } \right)^B \;
\left( {\dcal \over \Lambda } \right)^C . \eqn\strongvertexi $$

As a first application of this result consider the corrections
to the vector boson masses. For this take $ C+D=2, \ A = B = 0 $,
then $ M^2 \sim \Lambda^4 \lambda g^2 / \Lambda^2 = ( \Lambda g / 4 \pi )^2
= ( g \Lambda_ \phi )^2 $. From the expression for the vector boson masses
in a spontaneously broken gauge theory it follows that $ \Lambda_ \phi
= v $ where $v$ is to be identified with the \vev\ of the scalars
in the \sm. Then $$
\Lambda_\phi = v \simeq 246 \gev, \quad
\Lambda = 4 \pi v  \simeq 3 \tev \quad
\Lambda_\psi = ( 4 \pi )^{1/3} v \simeq 572 \gev \eqn\eq $$

For the operator $ \ocal_W $ I must take $ C + D = 6 $ and
$ A = B = 0 $ whence $$ \alpha_W \sim \inv{ \Lambda^2 }
{g^3 \over 16 \pi^2 } \eqn\eq $$ which is the same estimate
as was obtained in the decoupling case.

\section{{\it Many particles.}}

In this subsection I wish to consider the
situation where there are operators which are loop generated
(when the heavy physics is weakly interacting), but when there
are very many graphs contributing to such operators. In
particular, what would happen if there were $ \sim 160 $
graphs adding coherently in their contributions to the said
operator?

For the vector bosons we would expect corrections $ \sim 100 \% $
to their masses. For the scalar masses the contribution of $N$ graphs
of the type

\setbox2=\vbox to 2 truein{\epsfysize=4 truein\epsfbox[0 0 612 792]{figm10.ps}}
\centerline{\box2}
\vskip -50pt

\noindent(where the particles in the loop have a mass $ \sim
\Lambda $) will generate corrections
$ \sim N  \Lambda^2 / 16 \pi^2 $ to the scalar's mass.
Thus if there are $ \sim 160 $ such contributions adding coherently
the masses of these scalars become $ O ( \Lambda ) $; the
corresponding fields are in fact not a part of the low energy
lagrangian. If there are many contributions to the loop
generated operators then one should expect very strong deviations
from the naive low energy lagrangian: all (unprotected) scalar
fields disappear from $ \leff $ and the protected masses get
very large corrections.

\section{{\it Gauge invariance.}}

\REF\veltmaniruv{M. Veltman, {\it Acta Phys. Polon.}{\bf B12} (1981) 437.}
When considering effective theories it is often assumed that
the gauge invariance present in the \sm\ is in fact a low energy
symmetry broken at higher energies.
I will argue that this
assumption is inconsistent with the lightness of the gauge
boson masses and with the experimental result that the
gauge coupling constants are all small~\refmark{\veltmaniruv}.

Consider a theory with vector bosons and fermions;
the vector
bosons have a mass of order $M$ and
propagator $ ( g_{ \mu \nu } - p_\mu p_\nu /M^2 )
/ ( p^2 - M^2 +i \epsilon ) $. I now consider
consider the radiative
corrections to the triple vector boson couplings and to the
fermion-anti fermion-gauge boson couplings.
The vertices of interest are

\setbox2=\vbox to 2 truein{\epsfxsize=3 truein\epsfbox[0 0 612 792]{figm11.ps}}
\centerline{ \box2 }

\vskip -30pt

\noindent where the factor of $p$ denotes a derivative (I will not need
a more precise description of the vertices). I then obtain
the following estimates

\setbox1=\vbox to 3 truein{\hsize 2 in  \vskip -1.4 in $$  \sim
                         \inv{ 16 \pi^2 M^4 } g_A^2 \Lambda^6
                         \sim M^2 $$ \par \vskip -2.3 in $$
                         \sim \inv{ 16 \pi^2 M^4 } g_f^3 \Lambda^2
                         \sim g_f $$ }
\setbox2=\vbox to 3 truein  {\epsfysize=4 truein\epsfbox[0 0 612
792]{figm12.ps}}
\line{\box2 \kern -1.15 in\box1}
\vskip -50pt

{}From these results it follows that
 $$ { g_A \over 4 \pi } \sim { M^3
\over \Lambda^3 } , \qquad { g_f \over 4\pi }
\sim { M \over \Lambda } . \eqn\eq $$
Thus, if as is the case for the \sm, $ g_A \sim g_f $, then $$
M \sim \Lambda ; \qquad g_A, g_f \sim 4 \pi \eqn\eq $$ This
implies that radiative corrections to the gauge
boson masses will shift them out of the low energy theory;
moreover, the corrections to the coupling constants are so big
as to render the theory strongly interacting.

This argument strongly supports the claim that gauge invariance cannot
be broken ``softly'' by higher dimensional operators.

\section{{\it Blind directions.}}

\REF\derujula{A. De R\'ujula \etal, \npb384 (1992) 3.}
Another consideration relevant for quantitative estimates is the
possible presence of
``blind directions''~\refmark{\derujula}: these are operators
to which we have no experimental sensitivity since they affect
quantities which are precisely measured only at the one loop
level (or beyond). An example is the operator
$$ \ocal_W = \epsilon_{ I J K  } W_{ \mu \nu }^I W_{\nu \lambda }^J
W_{ \lambda \mu }^K , \eqn\ow $$ which is to be contrasted with
the operator $ \ocal_{ W B } = \left( \phi^\dagger \sigma_I \phi \right)
W_{ \mu \nu }^I B^{ \mu \nu } $ which contributes to the oblique
$S$ parameter and is well measured.

It has been assumed that the coefficients of these two operators
are of the same order as it appears difficult to suppress one
with respect to the other without fine tuning. This is in fact not
the case: it is easy to generated models where there is such a natural
suppression, so that the assumption that blind directions can
be estimated based on the ``sighed'' directions remains an additional
assumption to be tested by experiment.

Consider first the following toy model
consisting of a light
scalar field $ \phi $ interacting with two heavy fermions $ \psi_a \ (a = 1 ,2
)$. The lagrangian is $$ \lcal = \underbrace{ \half ( \partial \phi )^2 - \half
m^2 \phi^2 - \inv6 \sigma \phi^3 - \inv{24} \lambda \phi^4 }_{\hbox{{\sanser
light \ sector }}} + \underbrace{ \sum_{ a = 1 }^2 \bar \psi_a
\left( i \not \!
\partial - M + ( - )^a g \phi \right ) \psi }_ {\hbox{{\sanser heavy \ sector
}}} \eqn\eq $$ the Feynman rules for this model are

\setbox2=\vbox to 2 truein{\epsfysize=5 truein\epsfbox[0 0 612 792]{figm13.ps}}
\centerline{\box2}
\vskip -50pt

\noindent from which the graphs

\setbox2=\vbox to 2 truein{\epsfxsize=3 truein\epsfbox[0 0 612 792]{figm14.ps}}
\centerline{\box2}
\vskip -50pt

\noindent are seen to
have a coefficient $ \left[ \left( -1 \right) ^a g \right]^n $.
Since the contribution to the $n$ scalar point function is the
sum over $a = 1 ,2 $ it is clear that the contributions with an odd
number of scalar legs cancel.

If low energy ($ \ll M $)
experiments are sensitive to, for example, $ \phi^5 $, but
not to $ \phi^6 $ (which is then a blind direction),
there would be no experimental indication of the heavy sector.
The null results could be interpreted as
$ \Lambda $ being astronomical raising strong objections for
spending any resources in
measuring the effects of $ \phi^6 $,
while in fact a new generation of
experiments could very well uncover the presence of the heavy fermions.

The same results can be obtained in the \sm\
by considering two vector-like fermions
of almost degenerated masses with opposite hypercharges.
Then there are no contributions generated to any
operator with an odd number of $B$ legs, in particular
$ \ocal_{ W B } $ is not generated, while the contributions
to $ \ocal_W $ from the fermions add up and so the coefficient
is $ \sim g^3 /16 \pi^2 $

\section{{\it Equations of motion and orders of magnitude.}}

It is easy to find situations where two operators are equivalent in the
sense that $$ \ocal - \ocal' =
\int d^4x \; \acal ( \phi ) { \delta S \over \delta \phi } \eqn\eq $$
where $ \ocal $ is tree level generated but $ \ocal' $ is
generated at the one loop level. It is then quantitatively important
{\it not} to eliminate $ \ocal $ in favor of $ \ocal' $ since the
ensuing estimates for the coefficients can be wrong by several orders
of magnitude.

\chapter{Effective lagrangians at one loop.}

I will consider now the one loop contributions generated
by effective lagrangians. In many situations these contributions are
negligible. This is usually so whenever the effective operator
in a graph is loop generated: such graphs are in fact
the low energy limit of certain two loop diagrams in the full theory.

There are however situations in which the effective-operator one loop
contributions are important. Firstly when considering high precision
data involving tree-level effective operators (whose coefficients do not have
a loop suppression factor).
But other situations can be imagined, for example it is possible for
certain processes to be forbidden at tree level (with or without the
presence of effective operators) and for the loop contributions
without effective operators to be suppressed. In
this case the loops involving effective operators can be of
importance.

\REF\gunionetal{J.F. Gunion \etal, {\it The Higgs Hunter's Guide},
                (Addison-Wesley, Redwood City, CA, 1990).}
An example of this last case is the decays of the CP-odd scalar $ a_o $
present in the two scalar doublet extension of the \sm~\refmark{\gunionetal}.
I assume flavor changing neutral currents are suppressed by imposing
a discrete symmetry,
and consider the process $ a_o \rightarrow \gamma \gamma $
generated by (only) quark loops.

\setbox2=\vbox to 2 truein {\epsfysize=4.5 truein \epsfbox[0 0 612
792]{figm15.ps}}
\centerline{\box2}
\vskip -50pt

\REF\perez{M.A. Perez \etal, in preparation.}
\noindent Since the couplings are\foot{{\ninerm $ \tan \beta $
denote the ration of the \vev s of the scalar doublets in this model.}}
$ a_o \bar t t
\propto m \lowti{top} / \tan\beta $ and $ a_o \bar b b
\propto m \lowti{bottom} \tan \beta $; then for $ \tan \beta \gg 1 $,
$ B ( a_o \rightarrow \gamma \gamma ) \simeq 0 $.
This suggest we study the effects of higher dimensional
operators for this model (generated by physics beyond
the two doublet model)~\refmark{\perez}.
When this is done it is easy to verify that
the discrete symmetries forbids all $ a_o \gamma \gamma $ couplings
in operators of dimension $ \le 6 $.
Since there are no tree level contributions the loop graphs will be
finite, as we will see. Moreover these loop contributions
need not be small in the $ \tan\beta \gg 1 $ limit\foot{{\ninerm
The effective lagrangian contributions in this example can also
dominate in the $ a_o $ creation in photon-photon collisions.}}

Finally there is one last reason why loop calculations are important: I
have stated above that effective theories are actually renormalizable
and the verification of this claim with explicit computations is of
importance.

\section{{\it Dimensional regularization.}}

I will present here a very brief introduction to dimensional
regularization; for a thorough review see Ref.~\collins.
I will limit myself to zero spin bosonic theories (though no new
concepts are involved when this is extended to vector bosons
and fermions, the only
complication concerns the proper definition
of $ \gamma_5 $). As mentioned previously the
idea is simply to define everything in $n$ dimensions and letting $ n
= 4 - \epsilon $ with the understanding that $ \epsilon \rightarrow 0 $
at the end of the computation.

The first ingredient is Feynman's trick for combining denominators:

$$ \prod_i \inv{ A_i ^{ \alpha_i } }
= { \Gamma \left( \sum_i \alpha_i \right)
\over \prod_i \Gamma ( \alpha_i ) } \int_0^1 \prod_i d x_i \;
\delta \left( 1 - \sum_i \alpha_i \right)
{ \prod_i x_i^{ \alpha_i -1 } \over \left( \sum_i A_i x_i \right)^{ \sum_i
\alpha_i } } \eqn\eq $$ where all the $ A_i $ are assumed to have the
same-sign imaginary parts.
For example, $ 1/ (  A_1 A_2 ) = \int_0^1  dx \; \left[ x A_1 + ( 1 - x )
A_2 \right]^{ -2 } $.

As the second ingredient I need the Feynman rules. To avoid
unnecessary complications I will consider a simple
theory whose lagrangian is $$ \lcal = \half \left( \partial \phi \right)^2
- \half m^2 \phi^2 - \inv{ 4! } \lambda \phi^4 \eqn\eq $$ then the
Feynman rules are

\setbox1=\vbox to 30 pc{\hsize 2 in \vskip -28 pc \noindent $$
                         { i \over p^2 -m^2 + i \epsilon }
                         $$ \par \vskip -46 pc \noindent $$
                         - i \lambda  $$ }
\setbox2=\vbox to 30 pc{\epsfysize=30pc\epsfbox[0 0 612 792]{figm16.ps}}
\line{\box2 \kern-14pc \box1}
\vskip -245pt

The theory is regulated by defining it in $n$ dimensions (the metric
has the form diag$ ( 1 , -1 , -1 , \ldots ) $.)
The recipe to generate the Green's functions in a (regulated) perturbative
expansion is to draw all
possible graphs, each vertex and line given as above
with momentum conservation imposed in each vertex.
One finds that some momenta are
not fixed by this condition, these are called
loop momenta. Any such loop momentum $ l $ must be
integrated over with measure $ d^n l / ( 2 \pi )^n $. Finally,
to each graph one has to associate
a symmetry factor $ 1 / w! $ for each group of $w$ lines
which can be permuted without altering the graph in any way.
The number of vertices determines the power of $ \lambda $ and,
therefore, the order in perturbation to which we are working.

For example, the diagram

\setbox1=\vbox to 2 truein{\epsfysize=4.5truein\epsfbox[0 -50 612
742]{figm17.ps}}
\centerline{\box1}
\vskip -50pt

\noindent (with the slashes in the external legs indicating that
the corresponding propagators are omitted)
corresponds to  the following integral $$ I_n
= \half ( - i \lambda ) \int { d^n k
\over ( 2 \pi )^n } { i \over ( k + p )^2 - m^2 + i 0 } { i \over k^2
- m^2 + i 0 } \eqn\eq $$ imagining for the moment that $n$ is an integer.
Next I combine denominators to obtain (I define $ d_nk =
d^nk / ( 2\pi )^n $)
$$ \eqalign{ I_n
& = \half ( + i \lambda ) \int d_n k \int_0^x dx \inv{ \left\{
x \left[ ( k + p )^2 - m^2 _ i 0
\right] + ( 1 - x ) \left( k^2 -m^2 +i 0 \right) \right\}^2 } \cr
&= { i \lambda \over 2 } \int d_n k \int_0^1 dx \;
\inv{ \left( k^2 + 2 x k \cdot p + x p^2 - m^2 + i 0 \right)^2 } \cr
&= { i \lambda \over 2 } \int d_n k \int_0^1 dx \;
\inv{ \left[ ( k + x p )^2 - \xi \right]^2 } ; \qquad \xi = m^2 -x (1 - x ) p^2
- i 0 .\cr
&={ i \lambda \over 2 } \int   d_n k \int_0^1 dx \; \inv{ \left( k^2 - \xi
\right)^2}
= { i \lambda \over 2 } \int_E d_n k \int_0^1 dx \; \inv{ \left( k^2 + \xi
\right)^2}
\cr } \eqn\eq $$ where in the last equalities I have
assumed that the integral is well defined,
exchanged order of integration, and shifted $ k \rightarrow k - x p $; I
then performed a Euclidean rotation, replacing $ k^0 \rightarrow i k^0 $
(the subscript $E$, which I will henceforth
drop, indicates that the integral is over Euclidean space).

The next step is to write the volume element in
terms of polar coordinates $ d_n k = k^{ n - 1 }
dk \; d \Omega_n / ( 2 \pi )^n $ where $ d \Omega_n $ is the element of
solid angle
in $n$ dimensions. In practically all applications the only property
of $ d \Omega_n $ which is needed is it integral over the whole of the
sphere in $n$ dimensions: $ \int d \Omega_n = 2 \pi^{ n/2 } /
\Gamma( n/2 ) $.
Then I get $$ I_n
=  { i \lambda  \pi^{ n /2 } \over ( 2
\pi )^n \Gamma ( n /2 ) } \int_0^\infty d k  \int_0^1 dx \;
{ k^{ n -1 } \over  \left( k^2 + \xi \right) }
= { i \lambda \over 2 } {  \Gamma ( 2 - n/2 ) \pi^{ n /2 } \over ( 2
\pi )^n } \int_0^1 dx\; \xi^{ ( n-4)/2 } \eqn\eq $$
In this expression $n$ is understood to be an arbitrary complex number,
the only restriction being that the integral is well defined;
for a discussion on the validity of this assumption see Ref. \collins.

The final step is to replace $ n = 4 - 2 \epsilon $ with the
understanding that $ \epsilon \rightarrow 0 $. Using $ \Gamma (
\epsilon ) = \inv \epsilon - \gamma_E + O ( \epsilon ) $ where
$ \gamma_E $ denotes Euler's constant, yields
$$  I_n = { i \lambda \over 32 \pi^2 }
\int_0^1 dx\; \left\{
\underbrace{ \inv \epsilon - \gamma_E + \ln \left( 4 \pi \right) }_{ =
\cuv } - \ln \xi + O ( \epsilon ) \right\} . \eqn\eq $$

The presence of a logarithm of a dimension-full quantity might seem a
puzzling result of the method. Its origin can be traced back to the
fact that in $n$ dimensions $ \lambda $ is not dimensionless: the
kinetic energy of the scalars fixes the canonical dimension of the
field at $ n/2 - 1 $, so that $ d^n x \; \phi^4 $ has dimensions $ n
- 4 = - 2 \epsilon $. It is therefore convenient to replace $ \lambda
\rightarrow \lambda \kappa^{ 2 \epsilon } $ where $ \kappa $ is an
arbitrary scale and $ \lambda $ is now dimensionless. The
arbitrariness in the choice of $ \kappa $ is far from being a defect
of the theory or the regularization method: the dependence of Green's
functions on $ \kappa $ determines the renormalization group flow of the
effective couplings; all observables are independent of this scale.
With this replacement I get
$$  I_n = { i \lambda \over 32 \pi^2 }
\left\{ \cuv - \int_0^1 dx\;
\ln \left( \xi / \kappa^2 \right) + O ( \epsilon ) \right\} . \eqn\eq
$$

In a similar manner one can derive $$
\jcal_{ k m } = \int { d^n \ell \over ( 2 \pi )^n } \;
{ \left( \ell^2 \right)^k \over \left( \ell^2 - \xi \right)^m }
= { (-)^{ k + m } i \over ( 4 \pi ) ^{ n/2 } } \xi^{ k - m + n/2 }
{ \Gamma \left( k +{ n \over 2 } \right)
\Gamma \left( m -  k - { n \over 2 } \right)  \over
\Gamma( m ) } \eqn\eq $$ It can be verified that $ ( \lambda / 2 )
\jcal_{ 0 2 } = I_n $.

In dimensional regularization we also have  $$
\int { d^n k \over \left( k^2 \right)^\ell } = 0 \eqn\dimdelta $$ for all
values of $ \ell $ and $n$. This is a definition whose justification
can be found in many of the excellent reviews on the
subject~\refmark{\collins}.

Similarly I can evaluate the graph

\setbox2=\vbox to 25 pc{\epsfysize=30pc\epsfbox[0 -500 612 292]{figm18.ps}}
\centerline{\box2}
\vskip -192pt

\noindent and obtain $$
\half ( - i \lambda )\int { d^n \ell \over
( 2 \pi )^n } \; { i \over \ell - m^2 + i \epsilon } = { \lambda
\over 2 } \jcal_{ 0 1 }  = { i \lambda \over 32 \pi^2 } m^2 \left[ \cuv + 1 +
\ln m^2 \right]  \eqn\masscorrection . $$
Note that this results is proportional to $ m^2 $;
had I used a cutoff  terms $
\propto \Lambda^2_{ UV } $ would also appear.

As advertised the results have divergences appearing as poles in
$ \epsilon = ( 4 - n ) /2 $. The fact that the corrections to the
mass in \masscorrection\ are $ \propto m^2 $ is a result of
\dimdelta. Collecting all one loop contributions I finally get

\setbox1=\vbox to 2.5truein{\hsize 2 in \vskip .5truein \noindent $$ =
                         { 3 i \lambda^2 \over 32 \pi ^2 } \cuv +
                         \hbox{finite}
                         $$ \par \vskip 1truein \noindent $$ =
                         { i \lambda \over 32 \pi^2 } m^2 +
                         \hbox{finite}  $$ }
\setbox2=\vbox to 2.5truein{\epsfysize=30pc\epsfbox[0 0 612 792]{figm19.ps}}
\line{\box2 \kern-14pc \box1}
\vskip 30pt

With these preliminaries I can give a bird's eye view of the renormalization
program. Consider the corrections to the propagator. It is clear that
we can separate graph as those that can be separated by cutting one
internal line (called one particle reducible or 1PR graphs)
and those that cannot (called one particle irreducible or
1PI graphs). Denoting the sum of all 1PI graphs by a grey disk, which I
define to be the object $ i \Pi $ below,
I have the following relation

\setbox1=\vbox to 1.5truein{\hsize 7.4 in \vskip .43truein \noindent
                              $$ i \Pi
                             $$ \par \vskip .05truein \noindent
                             $$ \qquad \qquad \qquad \qquad \qquad
\qquad \qquad \qquad \qquad \qquad \qquad \qquad
\qquad \qquad \qquad \qquad \qquad \qquad
                               = { i \over p^2 - m^2 + \Pi } $$ }
\setbox2=\vbox to 1.5truein{\epsfysize=30pc\epsfbox[0 0 612 792]{figm20.ps}}
\line{\box2 \kern -6.2truein \box1}
\vskip 30pt

\noindent and this implies that the physical mass $ \mp $
is determined by the relation $$ \mp = m^2 - \Pi ( p^2 = \mp^2 ) \eqn\eq $$
so that in the vicinity of $ \mp $,
$$ \inv{ p^2 - m^2 + \Pi } \simeq { Z \over p^2 - \mp^2 } ;
\qquad p^2 \simeq \mp^2 \eqn\defofz $$

Similarly we have

\setbox1=\vbox to 2.5truein{\hsize 2 in \vskip -1truein \noindent $$ =
                              i \Lambda  $$ }
\setbox2=\vbox to 2.5truein{\epsfxsize=3.2truein\epsfbox[0 0 612
792]{figm21.ps}}
\line{\box2\kern6pc \box1}
\vskip -50pt

I define the physical coupling constant as $ \lp = \Lambda ( \bar p ,
\bar q ) $ where $ \bar p = \bar q = ( 2m , 0 , 0 , 0 ) $

To one loop I get
$$ \eqalign{  \lp &= \lambda
- { 3 \lambda^2 \over 32 \pi^2 } \left[ \cuv + 2 - \ln { m^2 \over
\kappa^2 } \right] \cr
\mp &= m^2 \left[ 1 + { \lambda \over 32 \pi^2 } \left( \cuv + 1
- \ln { m^2 \over \kappa^2 } \right) \right] . \cr } \eqn\eq $$
Taking $ \lp$ and $ \mp $ as the input data the lagrangian parameters
$$ \eqalign{ m^2 &= \mp^2 \left[ 1 -
{ \lp \over 32 \pi^2 } \left( \cuv+ 1
- \ln { \mp^2 \over \kappa^2 } \right) \right] \cr
\lambda &= \lp \left[ 1
+ { 3 \lp^2 \over 32 \pi^2 } \left( \cuv + 2 - \ln { \mp^2 \over
\kappa^2 } \right) \right] \cr } \eqn\eq $$ and the important property of
renormalization theory is that with these replacements, and with
an appropriate rescaling of the field $ \phi \rightarrow Z^{ - 1/2 }
\phi $, all infinities disappear (this is true to all orders in perturbation
theory).

It is worth stopping for a minute at this point and determine a
simple way of obtaining the relationship between the $ 1 / \epsilon $
poles and the divergences in terms of the cutoff $ \Lambda $. To do this
note the estimate
$$ \int { d^n \ell \over ( 2 \pi )^n } \inv{ ( \ell^2 - m^2 )^k }
\propto \left( m^2 \right)^{ -k + n/2 } \eqn\eq $$ in dimensional
regularization. With an UV cutoff in contrast this integral
is $ \propto \Lambda^{ n - 2 k } $ if $ n > 2 k $, so
that all but the logarithmic divergences ``disappear'' under
dimensional regularization (which fact can be traced back to
the property \dimdelta). Note however that a quadratic divergence
when $ n = 4 $ becomes a logarithmic divergence when $ n = 2 $ so
that we can extract quadratic divergences from dimensional regularization
by assuming $ n = 2 - 2\epsilon $.

The four-point function has only a logarithmic divergence at one loop;
this corresponds to the $ 1 / \epsilon  $ pole (where $ 2 \epsilon
= 4- n $). The two point function has both quadratic and logarithmic
divergences, the latter was obtained above. For the former note that
when $ n = 2 - 2 \epsilon $  I have $$ \left. { \lambda \over 2 }
\jcal_{ 0 1 } \right|_{ n = 2 - 2\epsilon } = - { i \lambda \over
32 \pi^2 } \left( \cuv - \ln { m^2 \over \kappa^2 } \right) $$
In terms of an UV cutoff this same integral when $ n = 4 $ yields
$ - i \lambda \Lambda^2 / (  32 \pi^2 )
+ [ i \lambda / ( 32 \pi^2) ]  m^2  \; \ln ( \Lambda^2
/ m^2 ) $ so we can identify $ \cuv $ with
$ \Lambda^2 $ when $ n \simeq 2 $ (up to possible logarithmic and
constant corrections) and with $ \ln \Lambda^2 $ when $ n \simeq 4 $
(up to constant corrections).

The main advantage of dimensional regularization is that it preserves
many important symmetries, such as gauge invariance. It does present
a problem when confronted with chiral interactions which involve
$ \gamma_5 $ fermion couplings. The reason is that in four
dimensions $ \gamma_5 = \epsilon^{ \mu \nu \alpha \beta }
\gamma_\mu \gamma_\nu \gamma_\alpha \gamma_\beta $ (up to
a normalization constant) and the extension to $n$ dimensions is
problematic since the $ \epsilon $ tensor is peculiar to four-dimensional
space time. The prescription that works~\refmark{\collins} is to define $
\gamma_5 = \gamma_0 \gamma_1 \gamma_2 \gamma_3 $ which satisfies
$$ \{ \gamma_\mu , \gamma_5  \} = 0  \quad \mu = 0 , 1 , 2 , 3 ;
\qquad [ \gamma_\mu , \gamma_5 ] = 0 \quad \mu \not= 0 , 1 , 2 , 3 \eqn\eq $$

\section{{\it Example.}}

I will now apply the above formalism to consider the radiative corrections
in an effective theory. I will assume a weakly-coupled underlying
physics and, for the low energy theory, I will take the
same simple scalar theory with a set of higher dimensional
operators added to it. To simplify
matters I will assume a symmetry under $ \phi \leftrightarrow - \phi $
so that the lowest dimensional operators have dimension six. The only
possibilities are then (symbolically)
$ \phi^2 \partial^4 , \ \phi^4 \partial^2 , \ \phi^6 $.

\item{ \phi^2 \partial^4 :} Two operators of this type are
$ \left( \partial_\mu \square \phi \right) \partial^\mu \phi $
or $ \left( \partial_\mu \partial_\nu \phi \right)^2 $, both of these
are equivalent (up to a total derivative) to the operator $ \left(
\square \phi \right)^2 $ which is the remaining possibility.
The classical equations of motion
read $ \square \phi + m^2 \phi + \lambda \phi^3 /6 = 0 $, so
that I can replace $$ \left( \square \phi \right)^2
\rightarrow m^4 \phi^2 + \third \lambda m^2 \phi^4 + { \lambda^2 \over 36}
\phi^6 \eqn\eq $$

\item{ \phi^2 \partial^2 :} One possibility is
 $ \phi^3 \square \phi$ which, again by the use of the equations of motion,
is equivalent to $ m^2 \phi^4 + \lambda \phi^6 / 6 $. The remaining
possibility is $ \phi^2 \left( \partial_\mu \phi \right)^2 $ which
is equivalent, via the equations of motion and integration by parts, to
$ m^2 \phi^4 /3 + \lambda \phi^6 / 18 $.

{}From these arguments it follows that, modulo a renormalization of
$m$ and $ \lambda $, the only operator I need to consider is $ \phi^6 $:
$$ \leff = \half \left( \partial \phi \right)^2 - \half m^2 \phi^2
- { \lambda \over 4! } \phi^4 + { \alpha \over \Lambda^2 } { \phi^6
\over 6! } , \eqn\eq $$  which is renormalizable to $ O ( \alpha ) $.

The only new divergences generated by the dimension six operator

\setbox1=\vbox to 2.5truein{\hsize 3 in \vskip .7truein \noindent $$ =
                              - i \left[ - { \alpha \over 32 \pi^2 }
                             { m^2 \over \Lambda^2 } \cuv +
                             \hbox{finite} \right] $$ \par \vskip .8 truein
                             $$ = - { i \alpha \lambda \over \Lambda^2 }
                             { \cuv \over 32 \pi^2 } + \hbox{finite} $$ }
\setbox2=\vbox to 2.5truein{\epsfxsize=5truein\epsfbox[0 0 612 792]{figm22.ps}}
\line{\box2\kern-22pc \box1}
\vskip 30pt

As in the previous case I can define the physical $ \phi^6 $ coupling
as

\setbox1=\vbox to 2.5truein{\hsize 2 in \vskip -.9truein \noindent $$ =
                              { i \over \Lambda^2 } \ap $$ }
\setbox2=\vbox to 2.5truein{\epsfxsize=5truein\epsfbox[0 0 612 792]{figm23.ps}}
\line{\box2\kern-14pc \box1}
\vskip -50pt

\noindent where all the space components of the external momenta are
assumed to be zero.

Now I can study the renormalization of the model. The input
parameters I choose are the physical
mass and couplings; these are defined by (I ignore higher loop corrections)
$$ \eqalign{
\mp^2 &= m^2 \left[ 1 + { \lambda \over 32 \pi^2 } \cuv + \hbox{finite}
\right] \cr
\lp &= \lambda - { 3 \lambda^2 \over 32\pi^2 } \cuv - { \alpha \over 32
\pi^2 } { m^2 \over \Lambda^2 } \cuv + \hbox{finite} \cr
\ap &= \alpha \left[ 1 - { 15 \lambda \over 32 \pi^2 } \cuv + \hbox{finite}
\right] \cr } \eqn\eq $$
which can be inverted to read
$$ \eqalign{
m^2 &= \mp^2 \left[ 1 - { \lp \over 32 \pi^2 } \cuv + \hbox{finite}
\right] \cr
\lambda &= \lp + { 3 \lp^2 \over 32\pi^2 } \cuv + { \ap \over 32
\pi^2 } { \mp^2 \over \Lambda^2 } \cuv + \hbox{finite} \cr
\alpha &= \ap \left[ 1 + { 15 \lp \over 32 \pi^2 } \cuv + \hbox{finite}
\right] \cr } \eqn\eq $$

As before it is easy to verify that
the whole of the renormalization program works: to any number
of loops (but keeping a single insertion of the dimension six operator!)
the replacement of Lagrangian parameters by their physical counterparts
renders all results finite. At two loops (and beyond) the only new effect is
the appearance of a divergence that is cancelled by an unobservable
rescaling of the
field: $ \phi \rightarrow \phi / \sqrt{ Z } $ for some constant $Z$,
defined in \defofz.

To $ O ( \alpha^2 ) $ the above statements are not valid. Consider for
example the graph

\setbox2=\vbox to 2.5truein{\epsfxsize=5truein\epsfbox[0 0 612 792]{figm24.ps}}
\line{\box2\kern-14pc \box1}
\vskip -50pt

\noindent is associated with the operator $ \phi^8 $; it diverges, and
it is of order $ \alpha^2 $. If one needs the $ O ( \alpha^2 ) $
corrections then
the operators of dimension eight should be included in the effective
lagrangian. Now the renormalization program can be carried out to
any loop order provided the graphs are restricted to having two
dimension-six operator insertions or one dimension-eight operator
insertion.

\section{{\it Gauge theories.}}

As discussed above the effective lagrangian must be required to be
gauge invariant. One can always chose the unitary gauge, but in this case
the propagator is, for a massive gauge boson (the mass
is understood to be a result of spontaneously symmetry breaking)  $$
{ - i \over p^2 - m^2 + i \epsilon } \left( g_{ \mu \nu }
- \inv{ m^2 } p_\mu p_\nu \right) \delta_{ a b } { \buildrel
p \rightarrow \infty \over \longrightarrow} { i \over m^2 }
{ p_\mu p_\nu \over p^2 } \delta_{ a b } \eqn\eq $$ where $p$ is
the momentum in the propagator and $a$ and $b$ are gauge indices
(if any). Since the unitary
gauge propagator does not vanish at infinite momentum the various
graphs are more divergent that in other gauges, such as the Feynman
gauge. Because of these complications this choice of gauge is not very
good for doing loop calculations; a more convenient
choice is the Feynman gauge propagator
$$
{ - i \over p^2 - m^2 + i \epsilon } \delta_{ a b } . \eqn\eq $$
In this case, however, we pay the price of having to include the
contributions of the unphysical scalars.

In the Feynman gauge one must also include the contributions from the
Fadeev-Popov ghost and the question naturally arises as to whether
I should also consider effective operators containing these fields. The
answer to this is no: effective operators are produced by integrating
out heavy degrees of freedom and are therefore independent of the
gauge fixing procedure used in the light sector of the model, hence
no dependence on the ghosts can appear in the effective operators.

\REF\bfg{B.S.DeWitt, in {\it Quantum Gravity 2}, edited by C.J. Isham,
R. Penrose and D.W. Sciama. G. 't Hooft, in {\it Karpacz 1975}, Acta
Universitatis Wratislaviensis, No 368, Vol 1. (Wroclaw 1976), pp
345. L.F. Abbot, {\it Act. Phys. Pol.} {\bf B13} (1982) 33.
M.B. Einhorn and J. Wudka, \prd39 (1989) 2758.}
Another remark is pertinent at this point. The effective action is
constructed by the following two step procedure. Consider the
full action $S$ and denote the light fields by $ \phi $ while the
heavy fields are denoted by $ \Phi $. Then construct $W$ using
 $$ e^{ i W } = \int [ d\phi ] [ d \Phi ]
e^{ i S + \int \phi j } \eqn\eq $$ where $j$ denotes the sources for the
light fields. Define then $ { \delta W \over \delta \phi } = \varphi
$ which is clearly the average of the light field $ \phi $
in the presence of sources. Finally define the effective action as
$$ \Gamma = W - \int j \varphi \eqn\eq $$ The gauge in $S$ can be
fixed so that $ \Gamma $ is a gauge invariant functional of the
light fields $ \varphi $~\refmark{\bfg}. Expanding $ \Gamma $ in powers of
$ 1 / \Lambda $ I obtain $ \Gamma = \int \leff d^4 x $. It follows
that $\leff $ is gauge invariant
and will not contain the ghost fields generated when fixing the
gauge for the light degrees of freedom.

\REF\arzti{C.Arzt \etal, U.C. Riverside report UCRHEP-T98 and
\prd\ (to appear).}
As a specific loop calculation involving a gauge theory consider
the contribution of $ \ocal_{ W B } $ to the anomalous
magnetic moment of the muon~\refmark{\arzti}. The relevant vertices are
$$ \eqalign{
\ocal_{ W B } = & \left( \phi^\dagger \sigma_I \phi \right) W_{ \mu \nu }^I
B^{ \mu \nu } \cr
\rightarrow & - 2 i g { v^2 \over 2 } \cw W_\mu^- W_\nu^+ F^{ \mu \nu }
+ \sqrt{8} \; \cw \phi_0 F^{ \mu \nu }
\left[ \phi_- \partial_\mu W_\nu^+ + \hbox{ herm. conj.} \right] \cr }
\eqn\eq $$ Its contribution appears in the following tree graphs;

\setbox2=\vbox to 2.5truein{\epsfxsize=4truein\epsfbox[0 0 612 792]{figm25.ps}}
\line{\box2\kern-14pc \box1}
\vskip -50pt

\noindent these yield $$ { m_\mu g \cw \alpha_{ W B } \over 16 \pi^2
\Lambda^2 } \left( \cuv + { 3 \over 2 } - \ln { \mw^2 \over \kappa ^2 }
\right) \sigma^{ \alpha \beta } k_\beta + \cdots \eqn\eq $$
which implies that $$ a_\mu =
{ m_\mu^2 \alpha_{ W B } \over 6 \pi^2 \tw
\Lambda^2 } \left( \cuv + { 3 \over 2 } - \ln { \mw^2 \over \kappa ^2 }
\right) \eqn\eq $$

The infinite contribution renormalizes the coefficients of the
effective operators
$ \bar \psi_\mu \sigma_{ \alpha \beta } \psi_\mu B^{ \alpha \beta }  $
and $ \bar \psi_\mu \sigma^{ \alpha \beta }
\sigma_I \psi_\mu W^I_{ \alpha \beta } $ which also generate a
tree-level contribution to $ a_\mu $~\refmark{\arzti}.

\chapter{Application potpourri}

When dealing with the practical applications of the effective
lagrangian formalism to electroweak processes there
are two cases of interest:
the decoupling case and the chiral case. In the first there is a
Higgs excitation in the light theory and the effective operators are
arranged according to their canonical dimension or, equivalently, to the
power of $ \Lambda $ in their prefactor. For the chiral case there are
no physical scalar excitations, and the terms in the effective lagrangian
are ordered according to the number of derivatives in the effective
operators.

For the decoupling case I will assume that the underlying physics is weakly
coupled. This is associated with the fact that it is very difficult to
maintain a light Higgs when the underlying physics
is strongly coupled; this either requires fine tuning or that
the low energy particle
content be radically altered~\refmark{\wudkarev}.

\section{{\it Four fermion operators.}}

There are two types of four-fermion operators which I will consider.
The first one is of the form $ ( \bar \psi_1 \gamma_\mu \psi_2 )
( \bar \psi_3 \gamma^\mu \psi_4 ) $, while the second
type is of the form $ ( \bar \psi_1 \psi_2 )
( \bar \psi_3  \psi_4 ) $. All other possibilities are equivalent via
Fierz transformations (all possible chiralities are understood
to be included). The coefficient of these operators can be
written in the form $ g_H^2 / \Lambda^2 $.

For the decoupling case these operators are generated by the graphs

\setbox2=\vbox to 2.5truein{\epsfxsize=4truein\epsfbox[0 0 612 792]{figm26.ps}}
\line{\box2 }
\vskip -50pt

\noindent where each vertex factor is $ g_H $ and the mass of the
exchanged particle is $ \Lambda $. A simple realization is the
low energy lagrangian for a model containing an extra neutral
$Z'$ vector boson. In this situation $ g_H \lesim 1 $.

For the chiral case the coefficients of the operators are estimated using
naive dimensional analysis. The same arguments used previously give
$ ( g_H / \Lambda )^2 \sim \Lambda ^4 \lambda / \Lambda_\psi^6 $
which, using $ \Lambda \sim ( 4 \pi )^{ 2/3 } \Lambda_\psi $
and $ 16 \pi^2 \lambda \sim 1 $, imply $$ g_H \sim 2 \sqrt{ \pi } \eqn\eq $$

In order to translate a limit for $ \Lambda $ obtained in the chiral case to a
limit
for the decoupling case one only needs to multiply  by $ 1/ \sqrt{ 4 \pi } \sim
0.3 $.
Thus in the decoupling case the sensitivity to new physics of any experiment is
significantly degraded.

\REF\ehlq{ E. Eichten \etal, {\it Rev. Mod. Phys.} { \bf56} (1984) 579.}
\REF\ruckl{R. R\"uckl, \plb129 (1983) 363; \npb234 (1984) 91.
           See also R.J. Cashmore \etal, {\it Phys. Rep.} 122 (1985) 275.}
For the chiral case LHC will be sensitive to scales below
$10$\tev~\refmark{\ehlq}
using the $ p_T $ distribution of jets. A better bound is obtained
by studying dilepton production~\refmark{\ehlq} $$ \Lambda \lesim 15 \tev
; \quad \hbox{ LHC \ ( chiral case) } \eqn\eq $$
A similar investigation for HERA yields~\refmark{\ruckl} $$ \Lambda \lesim 1
\tev
; \quad \hbox{ HERA \ (chiral  case) } \eqn\eq $$
For the decoupling case this last estimate translates into $ \Lambda
\lesim 300 \gev $ which is consistent with the previous sensitivity
limit obtained using helicity violating processes.

\section{{\it Triple vector-boson couplings.}}

\REF\hagiwara{ K. Hagiwara \etal, \npb282 (1987) 253.}
One set of effective couplings which have been extensively studied describe the
interactions between the $W$ bosons and the photon or the $Z$. These are
usually presented based on an effective lagrangian whose only explicit
symmetry is Lorentz and electromagnetic gauge invariances. The by
now standard representation is~\refmark{\hagiwara}
$$ \eqalign{
\lcal_{ WWV } / g_{ WWV } = &
i g_1^V \left( W_{ \mu \nu } ^\dagger W^\mu V^\nu - \hbox{ h.c. } \right)
- g_4^V
W_\mu^\dagger W_\nu \left( \partial^\mu V^\nu + \partial^\nu V^\mu
\right) \cr &
+ i \kappa_V W_\mu^\dagger W_\nu V^{ \mu \nu }
+ i { \lambda_V \over \mw^2 } W^\dagger_{ \lambda \mu }
W^\mu{}_\nu V^{ \nu \lambda } \cr
&  + i \tilde \kappa_V W^\dagger_\mu W_\nu \tilde V^{ \mu \nu }
+ i { \tilde \lambda_V \over \mw^2 } W^\dagger_{ \lambda \mu } W^\mu
{}_\nu \tilde V^{ \mu \nu } \cr
& + g_5^V \epsilon^{ \mu \nu \rho \sigma } \left(
 W_\mu {\buildrel \leftrightarrow \over \partial_\rho } W_\nu \right)
V_\sigma \cr } \eqn\tbv $$ (where terms
proportional to $ \partial \cdot W $ and $ \partial \cdot V $ are
ignored). $V$ denotes either the photon or the $Z$ field; in
the first case only terms in compliance with electromagnetic gauge
invariance are retained. It is assumed (without loss of generality)
that $ g_{ WW \gamma } = -e , \ g_{ WWZ }  = - e \cot \theta\lowti w $.
All field tensors are understood to be Abelian, for example,
$ V_{ \mu \nu } = \partial_\mu V_\nu - \partial_\nu V_\mu $.

For the chiral case the magnitudes of the coefficients are estimated using
naive
dimensional analysis. The results are
$$ \eqalign{
& | g_i^V | , | \kappa_V | , | \tilde \kappa_V | \sim { g^2 \over 16 \pi^2
} \sim 3 \times 10^{ - 3 } \cr
& | \lambda_V | , | \tilde \lambda_V | \sim { m_w^2 \over \Lambda^2 }
{ g^2 \over 16 \pi^2 } \sim 2 \times 10^{ - 6 } \left( { 4 \pi v
\over \Lambda } \right)^2 \cr } \eqn\eq $$
These estimates are modified should there be resonances
of masses $ \sim 4 \pi v $. In this case the presence of long tails could
enhance the above values by a factor of $ \lesim 10 $. This then
implies
$$ \eqalign{
& | g_i^V | , | \kappa_V | , | \tilde \kappa_V | \lesim 10^{ - 2 } \cr
& | \lambda_V | , | \tilde \lambda_V | \lesim 10^{ - 5 } \left( { 4 \pi v
\over \Lambda } \right)^2 \cr } \eqn\eq $$
Note also that for the chiral case  the $\lambda $ terms are of the form
$ ( \partial W )^3 $ corresponding to six covariant derivatives
(recall that a field tensor equals a commutator of covariant derivatives).
These terms are then subdominant.

For the decoupling case all coefficient are $ \sim ( v / 4 \pi \Lambda )^2 $
(with a possible enhancement of $ \lesim10 $).
It then follows that
implies
$$ | g_i^V | , | \kappa_V | , | \tilde \kappa_V | ,
| \lambda_V | , | \tilde \lambda_V | \lesim  \left( { 2 \times
10^{ - 3 } \over \Lambda^2\lowti{\tev} } \right)^2  \eqn\eq $$

Consider now the application of this lagrangian to tree level processes.

\REF\kim{C.S. Kim \etal, YUMS 93-15, SNUTP 93-44.}
\subsection{{\it HERA}}

In this case the best bound obtained is $ | \kappa -1 | \lesim 0.3
$~\refmark{\kim}
using the reaction $ e p \rightarrow \nu \gamma X $ and assuming five year's
integrated luminosity.

\REF\hagiwarazep{U. Baur and E.L. Berger, \prd41 (1990)
                 1476. K. Hagiwara \etal, \ibid\ pp 2113.}
\REF\bauri{U. Baur \etal, in
           proceedings of the {\it  Workshop on Physics at Current
           Accelerators and the Supercollider}, Aragonne, IL, 2-5 Jun 1993.}
\subsection{{\it Fermilab Tevatron}}

A similar investigation yields the constraints
$ | \kappa | , | \lambda | \lesim 2 $~\refmark{\hagiwarazep,\bauri}
using $ W^+ W^-$ production.

\subsection{{\it LEP2}}

For this accelerator a strong effort has been devoted to understanding
the properties of the reaction $e^+ e^- \rightarrow W^+ W^- $ when the
effects of \tbv\ are included. When the
cross section for this process is calculated, a term $ \propto
\lambda s^2 / \mw^2 $ is generated~\refmark{\hagiwara}.
This will generate
enormous deviations from the \sm\ prediction provided one (erroneously)
assumes $ \lambda \sim 1 $. If the natural order of magnitude
for $ \lambda $ is used all such effects
constitute small corrections to the \sm\ amplitude. This is
true for all effects derived from \tbv.

Not all corrections are negligible. Consider for example the presence
of a particle or particles that guarantee a unitary cross section.
Suppose also that the CM energy of the collider under consideration
is below the mass(es) of such particle(s). Then the unitarity
cancellations generated by these excitations are only partially
effective and we can get a glimpse at their presence. This in
fact is verified by explicit calculation for the corrections
to the
reaction  under consideration generated by a heavy fourth family.
The cross section acquires a correction factor of
$ \sim 1 + ( s/ \mw )^2 ( \mw/ 4 \pi v )^2 $ which can reach 10\%.
Note that it would be wrong to state that the corrections are
$ O ( s / \mw^2 ) $ since these are enormous; one must remember
that this factor is accompanied by a small coupling.

I would like to point out that
when calculating the reaction considered in this subsection it
is common practice to include only the diagrams generated
by a $Z$ or $ \gamma $ $s$-channel exchange.
There are however other graphs such as

\setbox2=\vbox to 2.5truein{\epsfxsize=4truein\epsfbox[0 0 612 792]{figm27.ps}}
\centerline{\box2 }
\vskip -50pt

\noindent generated by the \sm\ and by five dimension-six operators.
Finally there are contact terms
generated by the operator $$ i \left( \bar \ell \sigma_I
\gamma^\mu D^\nu \ell \right) W_{ \mu \nu } ^I = \half g \left(
\bar e_L \gamma^\mu e_L \right) W_{ \mu \nu }^- W^+{}^\nu + \cdots
\eqn\eq $$

If we assume that the order of magnitude of the neglected graphs
is similar to the one obtained from the contributions generated
by $ \lcal_{ WWV } $ then the graphs which are not included would
not alter the estimate on the sensitivity to the scale of new physics.
This however, constitutes an added assumption on the model and should
be kept in mind when evaluating the constraints on $ \Lambda $ derived
from various calculations.

\subsection{{\it LHC}}

\REF\boudjema{F. Boudjema, in {\it 2nd International Workshop on Physics
              and Experiments with Linear $ e^+ e^-$ Colliders},
              Waikoloa, HI, 26-30 Apr. 1993.
              (Bulletin Board: hep-ph@xxx.lanl.gov - 9308343).}%
\REF\falk{A.F. Falk \etal, \npb365 (1991) 523.}
\REF\barklow{T. Barklow, talk presented at
             the {\it 1993 Aspen Winter Conference on Elementary Particle
             Physics}, Aspen Center for Physics, January 10--16, 1993.}
Various investigations~\refmark{\boudjema,\falk,\barklow}
have considered limits on the coefficients of \tbv\ and
derive the sensitivity limits
$ | \kappa | $,$ | \lambda | \lesim 0.3 $
or $ | \kappa | $,$ | \lambda | \lesim 0.1 $
which are obtained
by considering final states with only two vector bosons,
$ W^+ W^-, \ ZZ, \ Z W^\pm, \ W^\pm W^\pm $. The most promising
of these is the $ZZ$ final state. The idea is then to measure
the $ p_T $ distribution of the $Z$ vectors and search for a
deviation from the \sm\ predictions.

\REF\bargerandphillips{J. Barger and R.J.N. Phillips, lectures
                       presented by J. Barger at the {\it VII Jorge
                       Andr\'e Swieca Summer School: Particles and
                       Fields}, Sao Paulo, Brazil, 10-23 Jan 1993.}
Note however that LHC will probe energies above $ 3 \tev $
where the non-linear realization of the symmetry (chiral case) is
not applicable. In order to deal with this problem one can
simply impose cuts that insure that this bound is
not violated~\refmark{\falk}; this of course degrades the signal to noise
ratio.
If no sever cuts are imposed, then a recipe for extending
the model to energies above $ 4 \pi v $ must be provided. This
is often done by postulating the appearance of new resonances
(see Ref.~\bargerandphillips\ and references therein).
Of these two possibilities the first one preserves all the
features of the effective lagrangian approach, the second one
is, by its very nature, model dependent.

\subsection{{\it NLC}}

For this accelerator, assuming $ \sqrt{s} = 1.5 \tev $
and a luminosity of $ 100 /$fb, the limits on the coefficients are
significantly
improved: $ | \kappa | \sim 10^{-2} - 10^{-3} $~\refmark{\barklow,\boudjema}.
For the chiral
case ton must
remember that this accelerator is close to the limit of validity
of the effective lagrangian parametrization, in fact $ \sqrt{s}
\sim (4 \pi v )/2 $ so that corrections of the order of
$ \sim (1/2)^2 \sim $25\%\ can be expected.
When $ \sqrt{s} = 500 \gev $ CM energy similar limits on $ \kappa $
are obtained.

For the chiral case one can also derive bounds on the
coefficients of $$ \lcal=
{ c_L g \over 16 \pi^2 } \tr\left\{ \WW_{ \mu \nu }
\left( D^\mu U \right)^\dagger \left( D^\nu U \right) \right\} +
{ c_R g' \over 16 \pi^2 } \tr\left\{ B_{ \mu \nu } \sigma_3
\left( D^\mu U \right)^\dagger \left( D^\nu U \right) \right\} \eqn\eq $$
where we expect $ | c_{ L,R} | \sim 1 $.
The results are~\refmark{\boudjema}

{\eightrm
\vbox{\tabskip=0pt \offinterlineskip
\def\tablerule{\noalign{\hrule}}
\halign to 5 in{\strut#& \vrule#\tabskip=1em plus 2em&
\hfil#& \vrule#& \hfil#\hfil& \vrule #&
\hfil#& \vrule#& \hfil#\hfil& \vrule #&\hfil#& \vrule#
\tabskip=0pt\cr\tablerule
\omit&height.1in&\multispan{9}&\cr
&&\multispan{9}\hfil Limits on $ | c_{ L , R } | $ \hfil&\cr
&&\multispan{9}\hfil (chiral case) \hfil&\cr
\omit&height .1in&\multispan{9}&\cr\tablerule
\omit&height.02in&\multispan{9}&\cr\tablerule
&&\omit\hidewidth Coupling\hidewidth&&
\multispan{7}\hfil Experiment (future) \hfil&\cr
\tablerule
&&\omit\hidewidth {}\hidewidth &&
\omit\hidewidth LEP2 \hidewidth&&
\omit\hidewidth NLC (1) \hidewidth&&
\omit\hidewidth NLC (2) \hidewidth&&
\omit\hidewidth LHC \hidewidth&\cr\tablerule
&& $ | c_L | $
 && 30 && 2 && 1 && 15 &\cr\tablerule
&& $ | c_R | $
 &&150 && 10 && 5 && 150 &\cr\tablerule}}}

\noindent where the sensitivity of other accelerators are included
for comparison.
The luminosity of LEP2 is assumed to be $ 500/$pb
and NLC (1) denotes a $500\gev $ machine at $ 10/$fb; NLC (2)
denotes a $ 1 \tev $ machine at $ 44/$fb. Finally, the LHC is
assumed to have a luminosity of $ 10/$fb. Note that a machine such
as the NLC will have the necessary sensitivity to probe the
region where the couplings have their natural sizes.

The sensitivity to $ \lambda $ is also calculated~\refmark{\boudjema}; the
result
being $ | \lambda | \lesim 0.01 $.

\REF\yehudai{E. Yehuday,\prd41 (1990) 33; \ibid\ {\bf D}44 (1991).}%
\REF\dawson{S. Dawson and G. Valencia Fermilab report FRMILAB-PUB-93/218-T
             and \prd\ (to appear).}
\subsection{{\it $ e \gamma $ and $ \gamma \gamma $ colliders.}}

When a laser photon is backscattered in an $ e^+ e^- $ collider
several observational windows are opened up.
The process $ e \gamma \rightarrow Z Z e $ can be used to measure $ \kappa $
to a precision of $ 0.1 $, similarly $ \gamma \gamma \rightarrow
WW, \ ZZ $ can also measure $ \kappa $ to a precision of $ 0.01
$~\refmark{\yehudai,
\dawson}.

\REF\cleo{E. Thorndike, CLEO collaboration, talk given at the
          {\it 1993 Meeting of the American Physical Society},
          Washington, D.C., April, 1993.}
\REF\hemackellar{X.-G. He and B. McKellar Univ. of Melbourne report
                 UM-P-93-52 (unpublished) (Bulletin Board:
                 hep-ph@xxx.lanl.gov - 9309228).}
\REF\peterson{K.A. Peterson, \plb282 (1992) 207.}
\subsection{$ b \rightarrow s \gamma $.}

This is a one loop process if flavor mixing is ignored in the dimension-six
operators. Form the CLEO bound~\refmark{\cleo} of
$ B ( B \rightarrow K^* \gamma ) = \left( 4.5 \pm 1.4 \pm 0.9 \right)
\times 10^{ - 5 } $ one can derive $ B ( b \rightarrow s \gamma )
< 5.4 \times 10^{ - 4 } $ which translates into $ | \kappa | ,
| \lambda | \lesim 10 $~\refmark{\hemackellar,\peterson,\dawson}.

\REF\marciano{T. Kinoshita and W. Marciano in {\it Quantum Electrodynamics},
              T. Kinoshita Ed. (World Scientific Singapore, 1990).}
\REF\ags{V.W. Hughes, AIP conference proceedings no. 187, 326 (1989.)
         M. May, AIP conference proceedings no. 176, 1168 (1988).}
\subsection{{\it AGS851}}

The one loop contributions stemming from $ \lcal_{ WWV } $
to the muon anomalous magnetic moment can be
calculated~\refmark{\marciano,\arzti}.
In terms of the effective operator parametrization used in these lectures,
the  only contributions come form $ \ocal_{ W B } $ and $ \ocal_W $.
Choosing the natural sizes for the coefficients the results are
$$ { g - 2 \over 2 } \sim \cases{
10^{-12} / \Lambda^2\lowti{Tev} & (decoupling case) \cr
10^{-13} & (chiral case) \cr } \eqn\eq $$ while the sensitivity of
AGS851 is $ 10^{ -10 } $~\refmark{\ags}: the one loop contributions are
completely
negligible; the corresponding bounds for the coefficients of \tbv\
are $ O ( 10 ) $. In contrast the tree level contributions generate a
(non-trivial) bound of $ \sim 700 \gev $ for $ \Lambda $~\refmark{\arzti}.

\section{{\it Other operators.}}

In this section I give two examples of limits set on $ \Lambda $ using
other operators. this list is far from exhaustive (for a more comprehensive
compilation see Ref.~\wudkarev).

\REF\holdomi{B. Holdom, \plb259 (1991) 329.}
\REF\roy{D. Choudhury \etal,  Tata Institute report TIFR-TH/93-08
         (unpublished).}
\subsection{{\it S , T and U .}}

I have mentioned previously that these are important parameters to measure.
For the chiral case we expect $ S , \ U  \sim 0.1 $. In contrast, for
the decoupling case, $ S \simeq 0.3/ \Lambda^2\lowti{\tev} $
and $ U \sim 0 $
if $ \Lambda \gesim 3 \tev $~\refmark{\holdomi,\appelquistwu,\roy,\wudkarev}

\REF\hernandez{P. Hern\'andez and F.J. Vegas,  \plb307 (1993) 116.}
\subsection{{\it $Z$ widths, $ \tau $ polarization, $ \nu $ cross sections.}}

The sensitivity of the blind operators
$$ \ocal_1 = B_{ \mu \nu } \left( D^\mu \phi \right)^\dagger
\left( D^\nu \phi \right) ; \qquad
\ocal_2 = W^I_{ \mu \nu } \left( D^\mu \phi \right)^\dagger \sigma_I
\left( D^\nu \phi \right) ; \eqn\eq $$
to these processes can be calculated~\refmark{\hernandez}. With a natural size
for
the coefficients, the sensitivity to $ \Lambda $ is quite poor:
$ \Lambda \gesim 20 \gev $ for LEP1
$ \Lambda \gesim 100 \gev $ for LEP2.

\section{{\it Various.}}

Heretofore I assumed the effective lagrangian to have the same symmetry
structure as the \sm. In this section I will describe two instances where
this assumption is modified.

\REF\gounarisi{G. Gounaris \etal, report PM 93/26 (unpublished).}
\subsection{ $ \su2_R $ }

Up to now the only symmetries imposed on the effective lagrangian
were those of the \sm. One can of course modify this. For example,
one can assume that the underlying physics
conserves $ \su2_R $~\refmark{\gounarisi},
the approximate symmetry of the scalar sector.
Even though the kind of reactions
where the new physics effects are most noticeable are quite peculiar
in this case, one finds no measurable effect on LEP2 once the
natural size for the coefficients is used.

\REF\frere{J.-M. Fr\`ere \etal, \plb292 (1992) 348.}
\subsection{{\it Extended groups.}}

One need not require the effective lagrangian to be invariant under the
same group as the \sm. One may assume, for example, the presence of
an extra $ \ui $ factor, and that this extended group is broken to
the \sm\ group at a scale relatively close to the Fermi scale~\refmark{\frere}.
Since the gauge group is larger, there are more constraints on the
operators; on the other hand, since the particle content is increased,
there are more operators that can be constructed. Of these opposing
tendencies the second one dominates: there will be many more operators
when the gauge group is increased.

In this scenario there is an additional gauge boson $ B' $ associated with
the new $ \ui $ factor,
and one can consider the effects of the operator
$$ \ocal_{ W B' } = { \epsilon \over v^2 }
B'_{ \mu \nu } W^I {}^{ \mu \nu }
\left( \phi^\dagger \sigma_I \phi \right)  \eqn\eq $$ on a variety of
LEP2 observables. Using $W$ pair production significant effects are
found for the choices $ \epsilon = - 0 .2 $,
$ m_{ Z'' } = 300 \gev $ and when
the new gauge coupling $ g'' $ is $ 0.067 $.
Note however that the natural size for $ \epsilon $ is $ 0.04 $
for which value the effects of this operator on LEP2 processes is
negligible.

\chapter{Conclusions}

\item{\bullet} Effective lagrangians constitute a coherent model and
process independent scheme in which to study all possible effects
from physics beyond the \sm. The whole approach is general and consistent.

\item{\bullet} It has been unfortunately the case
that the effective lagrangian
approach has been often misused. For example, it is not
uncommon to find striking claims on the sensitivity of a given
experiment to $ \Lambda $ only to discover later that the
calculations do not respect the restriction $ \sqrt{s} < \Lambda $.

\item{\bullet} When used consistently they provide good,
(though not spectacular) bounds on the
scale of new physics: $ \Lambda \gesim 500 \gev$ (from HERA
and the muon anomalous magnetic moment); and $ \Lambda \gesim 15 \tev $
from the four-fermi operator effects at LHC.

\item{\bullet} One can certainly use them in perturbative calculations.
In a consistent scheme, however, these radiative corrections are almost
always negligibly small (due to the present precision attained in the
experiments).

\item{\bullet} Given the untimely demise of the SSC and the probable
postponement of the LHC this is a very good field in which to work given
that direct evidence of new physics might be a decade (or longer) away.

\ack
The author would like to thank the organizers for the
invitation to this course. I also would like to thank M. Einhorn, C. Arzt,
M.-A. Perez and J. Toscano for very many illuminating discussions.
This work was supported in part through funds provided by the DoE and
the SSC Laboratory.

\refout

\bye